\documentclass[letterpaper,english,reprint, aps]{revtex4-1}
\usepackage[T1]{fontenc}
\usepackage[latin9]{inputenc}
\usepackage{xcolor}
\usepackage{verbatim}
\usepackage{amsmath}
\usepackage{amssymb}
\usepackage{bbm}

\makeatletter

\providecommand{\tabularnewline}{\\}

\usepackage{graphicx}
\usepackage{xcolor}
\usepackage[bookmarks=false,linkcolor=blue,urlcolor=blue,colorlinks,citecolor=blue]{hyperref}
\usepackage{natbib}

\makeatother

\usepackage{babel}
\usepackage{soul}
\setstcolor{red}

\begin{document}

\preprint{}

\title{Benefits of weak disorder in one dimensional topological superconductors}

\author{Arbel Haim$^{1}$ and Ady Stern$^{2}$}

\affiliation{$^{1}$Walter Burke Institute for Theoretical Physics and the Institute
	for Quantum Information and Matter, California Institute of Technology, Pasadena, CA 91125, USA\\
	\mbox{$^{2}$Department of Condensed Matter Physics, Weizmann Institute of Science, Rehovot 7610001, Israel}}

\begin{abstract}
	Majorana bound states are zero-energy modes localized at the ends
	of a one-dimensional (1D) topological superconductor. Introducing disorder usually
	increases the Majorana localization length, until eventually inducing
	a topological phase transition to a trivial phase. In this work we
	show that in some cases weak disorder causes the Majorana
	localization length to {\it decrease}, making the topological phase more
	robust. Increasing the disorder further eventually leads to a change
	of trend and to a phase transition to a trivial phase. Interestingly
	the transition occurs at $\xi_0\gg l$, where $l$
	is the disorder mean-free path and {\normalsize{}$\xi_0$}
	is the localization length in the clean limit.
	Our results are particularly relevant to a 1D topological superconductors formed in planar Josephson junctions.
\end{abstract}



\maketitle

\paragraph*{\label{sec:Intro}Introduction.\textemdash{}}

Understanding the effect of unavoidable disorder on topological
superconductivity is of great interest. Of particular interest is its effect on the localization length of the zero-energy
Majorana bound states (MBSs), and the critical strength for transition to a trivial state.

Effects of disorder on spinless single-channel \emph{p}-wave superconductor~\cite{Kitaev2001unpaired,Read2000paired} - the canonical model for topological superconductivity (TSC)~\cite{Qi2011topological,Alicea2012,Beenakker2013,Lutchyn2018majorana,Aguado2017majorana} -  were previously studied~\cite{Motrunich2001Griffiths,Brouwer2011Probability,Lobos2012interplay,Pientka2013signatures,Huse2013localization,Adagideli2014effects}. Disorder was found to
increase the Majorana localization length, $\xi$,
according to $1/\xi=1/\xi_{0}-1/2l$, with $\xi_{0}$ being the localization
length (or coherence length) in the clean limit, and $l$ being the impurity-induced mean
free path~\cite{Brouwer2011Probability}. At the critical value
$l_{{\rm c}}=\xi_{0}/2$ the localization length diverges leading to a phase transition to a trivial phase.
Accordingly, the critical mean-free time, $\tau_{{\rm c}}$, is determined by the excitation gap of the clean system,
$\tau_{{\rm c}}^{-1}=2E_{\rm gap}$~\footnote{This result is valid when the Fermi energy is large compared with $E_{\rm gap}$, which is the limit of interest here. For the opposite limit see Ref.~\cite{Pientka2013signatures}.}.

For a multi-channel 1D system \cite{Potter2010multichannel,Rieder2013reentrant,Rieder2014Density,Lu2016influence,Burset2017current}, at weak-enough disorder
the behavior is similar to the single-channel case with monotonically-increasing $\xi$. For stronger disorder, multiple transitions between trivial and topological occur
at $l_{{\rm c}}^{(n)}=n\xi_{0}/(N+1),$ with
$N$ the number of channels~\cite{Rieder2013reentrant,Rieder2014Density}.

\begin{figure}
	\begin{tabular}{lr}
		\hskip -5mm
		\includegraphics[clip,trim= 0 0 0 2mm,height=4cm]{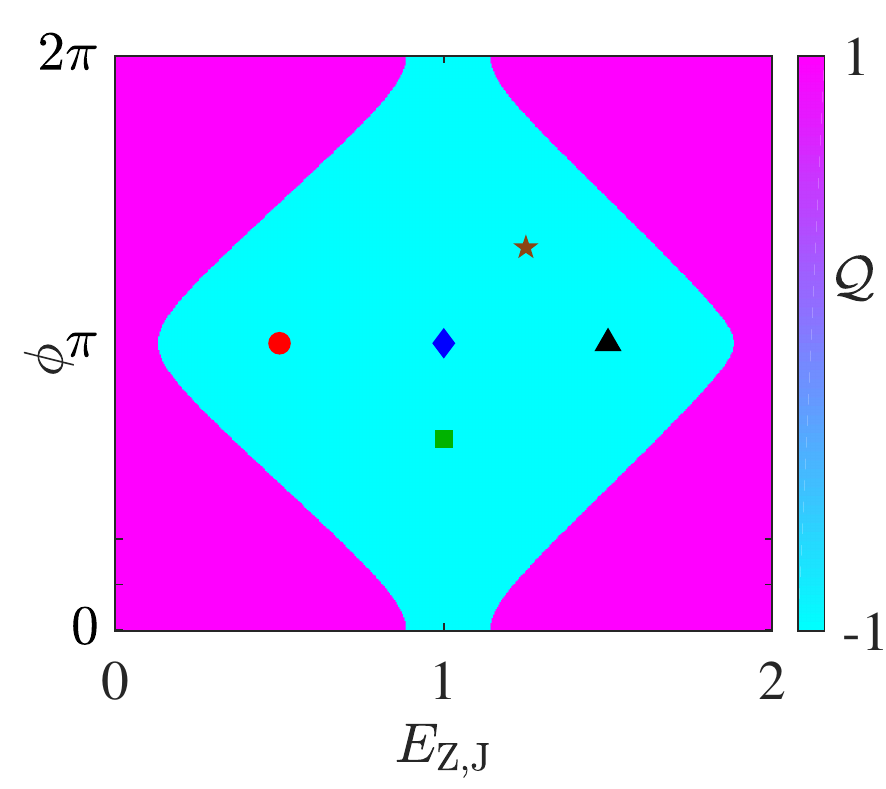}
		\llap{\hskip 2mm \parbox[c]{7.5cm}{\vspace{-3mm}(a)}}
		\hskip -2mm
		\includegraphics[clip,trim = 0mm 0mm 0mm 0mm,height=3.9cm]{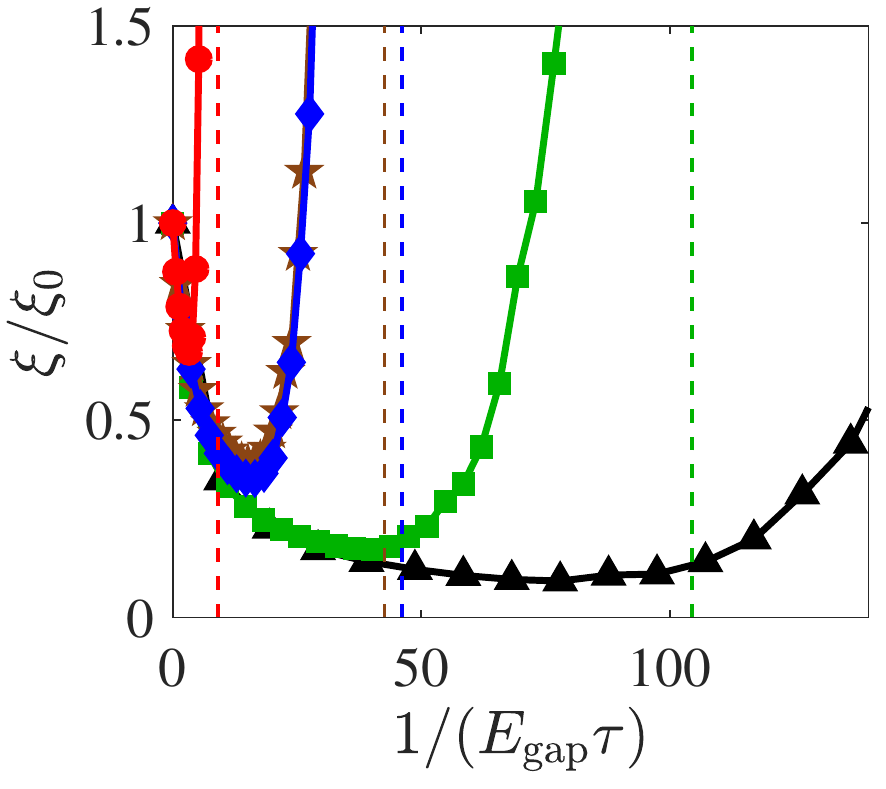}
		\llap{\hskip 2mm \parbox[c]{7.5cm}{\vspace{-3mm}(b)}}
		\hspace{6mm}{\llap{\includegraphics[clip=true,trim =0mm -8cm -7mm 0mm,height=3.3cm]{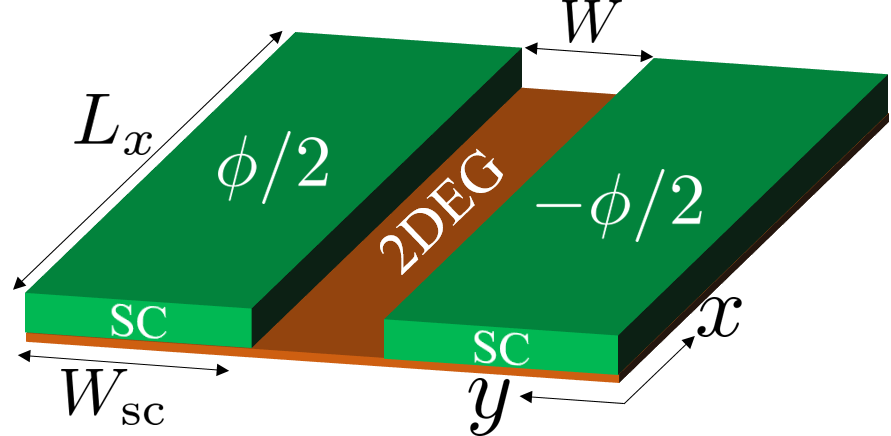}}}
	\end{tabular}
	\vskip -3mm
	\caption{(a) Phase diagram of the Planar Josephson junction, Eq.~\eqref{eq:H_PJJ}, in the clean limit. In the topological phase ($\mathcal{Q}=-1$), the system supports
		zero-energy Majorana bound states (MBSs) at each end of the junction. (b) The Majorana localization length, $\xi$,  versus the disorder-induced inverse mean free time, $\tau^{-1}$, for different points
		inside the topological phase [see markers in (a)]. At weak disorder,
		$\xi$ decreases with disorder. For stronger disorder $\xi$ increases
		until eventually diverging at the phase transition to the
		trivial phase. Here, $\xi$ is averaged over a 100 disorder realizations. The system parameters are $E_{{\rm so}}=m_{\rm e}\alpha^2/2=1$, $\Delta_0=1$, $\mu_{{\rm J}}=\mu_{{\rm SC}}=2.5$, $E_{\rm Z,SC}=0$, $l_{{\rm so}}=1/m_{\rm e}\alpha=0.2W$, $W_{\rm sc}=W$.
		In (b), $\tau$ is normalized by the overall gap in the clean system, $E_{\rm gap}$, which is $0.032$, $0.032$, $0.022$, $0.008$, and $0.004$ for the red, brown, blue, green and black plots, respectively. Similarly, $\xi$ is normalized by that of the clean system, $\xi_0$, which is $24W$, $34W$, $36W$, $93W$, and $194W$, in the same order.
		\label{fig:PJJ}}
\end{figure}

In this paper we study the effect of disorder on a novel realization of
a 1D topological superconductor: a planar Josephson junction (JJ),
implemented in a Rashba two-dimensional electron gas (2DEG), and subject
to in-plane magnetic field~\cite{Hell2017two,Pientka2017topological,Hell2017Coupling,Hart2017controlled} (see Fig.~\ref{fig:PJJ}). We find that in this system
weak potential disorder causes $\xi$ to \emph{decrease} [see. Fig.~\hyperref[fig:PJJ]{\ref{fig:PJJ}(b)}]. For strong disorder, the trend eventually reverses and the localization
length increases back until finally diverging at the transition
to the trivial phase. Importantly, this transition occurs at a critical
disorder strength, $\tau_{{\rm c}}^{-1},$ which is typically much
larger than the gap of the clean system.

\begin{figure}
	\begin{tabular}{lr}
		\rlap{\hskip -5mm \parbox[c]{0cm}{\vspace{-2.1cm}(a)}}
		\includegraphics[clip,scale=0.3]{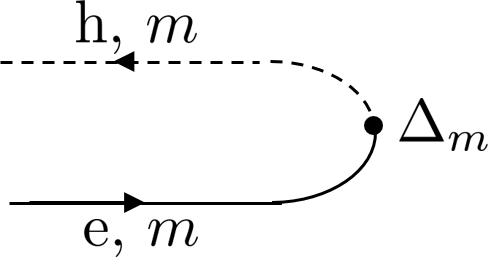}
		\hskip 5mm
		&
		\hskip 5mm
		\rlap{\hskip -5mm \parbox[c]{0cm}{\vspace{-2.1cm}(b)}}
		\includegraphics[clip,scale=0.3]{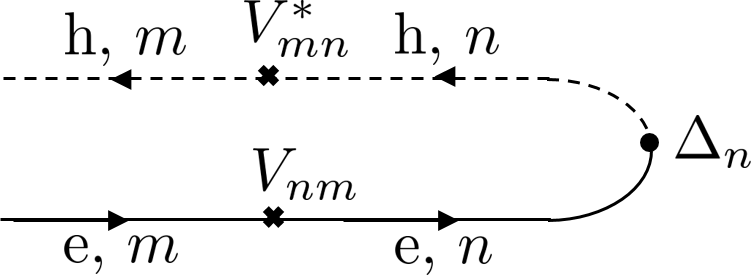}
	\end{tabular}
	\vskip -2mm
	\caption{
		(a) Andreev reflection (AR)
		in the $m$-th channel. (b) With disorder, the electron can
		first scatter to the $n$-th channel, perform AR,
		and then scatter back to the $m$-th channel.
		\label{fig:Low-energy-models}}
\end{figure}

Studying a general low-energy model for a multi-channel TSC, we show that disorder
can cause $\xi$ to increase or decrease,
depending on the relative phases of the pairing potentials in different channels, and the structure of the inter-channel impurity scattering (see also Fig.~\ref{fig:Low-energy-models}).
Scattering between modes of equal-phase pairing potential increases
the ``effective'' pairing gap, while scattering between
modes of opposite-phase potentials decreases the effective gap. Due to the \emph{p}-wave nature of the pairing within
each channel, intra-channel backscattering always decreases the effective
gap, and de-localizes the MBS.

\begin{figure}[h]
	\begin{tabular}{cc}
		\includegraphics[clip=true,trim =0mm 0mm 0mm 0mm,scale=0.43]{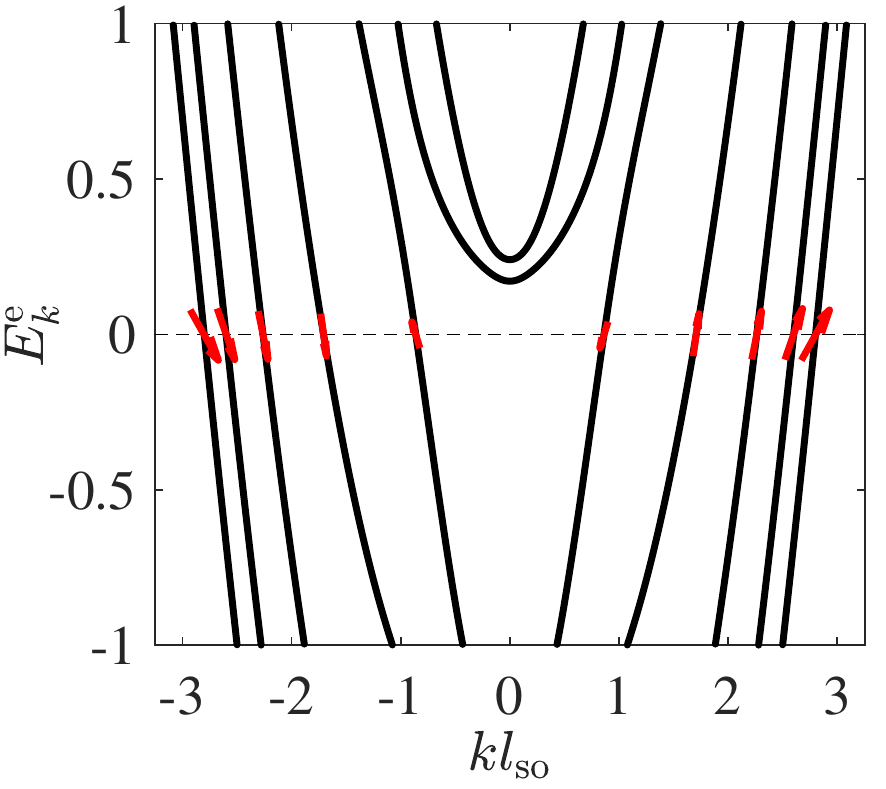}
		\llap{\hskip 2mm \parbox[c]{8cm}{\vspace{-6.3cm}(a)}}
		&
		\includegraphics[clip=true,trim = 1mm 0mm 0mm 0mm,scale=0.43]{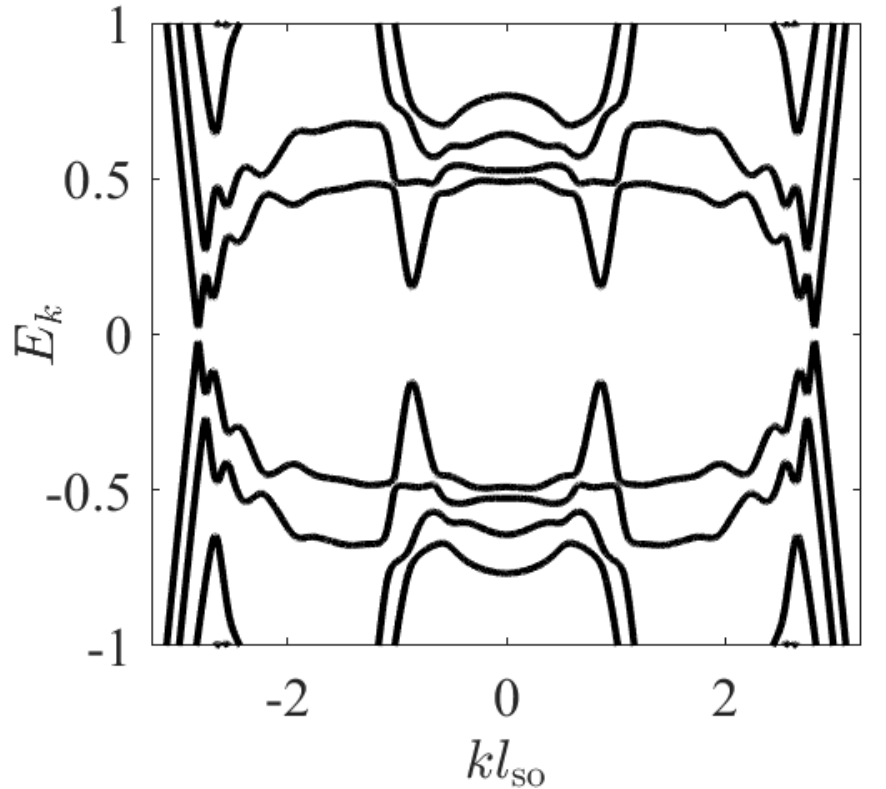}
		\llap{\hskip 2mm \parbox[c]{8cm}{\vspace{-6.3cm}(b)}}
		\vspace{-1mm}
		\tabularnewline
		\hskip 1.5mm
		\includegraphics[clip=true,trim =0mm 0mm 0mm 0mm,scale=0.45]{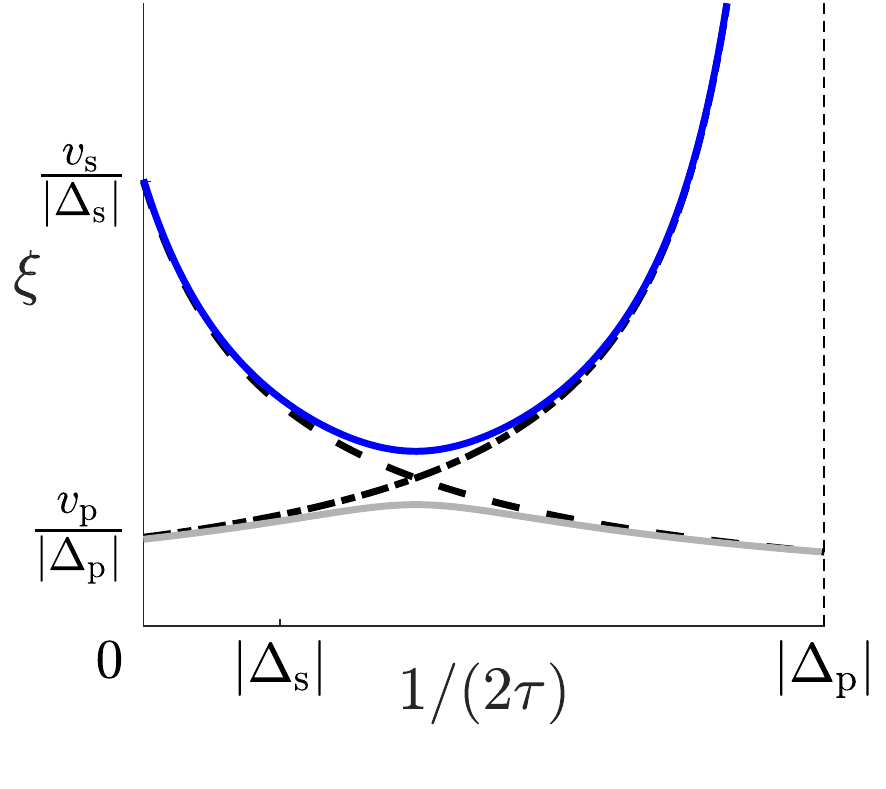}
		\llap{\hskip 2mm \parbox[c]{8cm}{\vspace{-6.7cm}(c)}}
		&
		\hskip 2mm
		\includegraphics[clip=true,trim =0mm 0mm 0mm 0mm,scale=0.45]{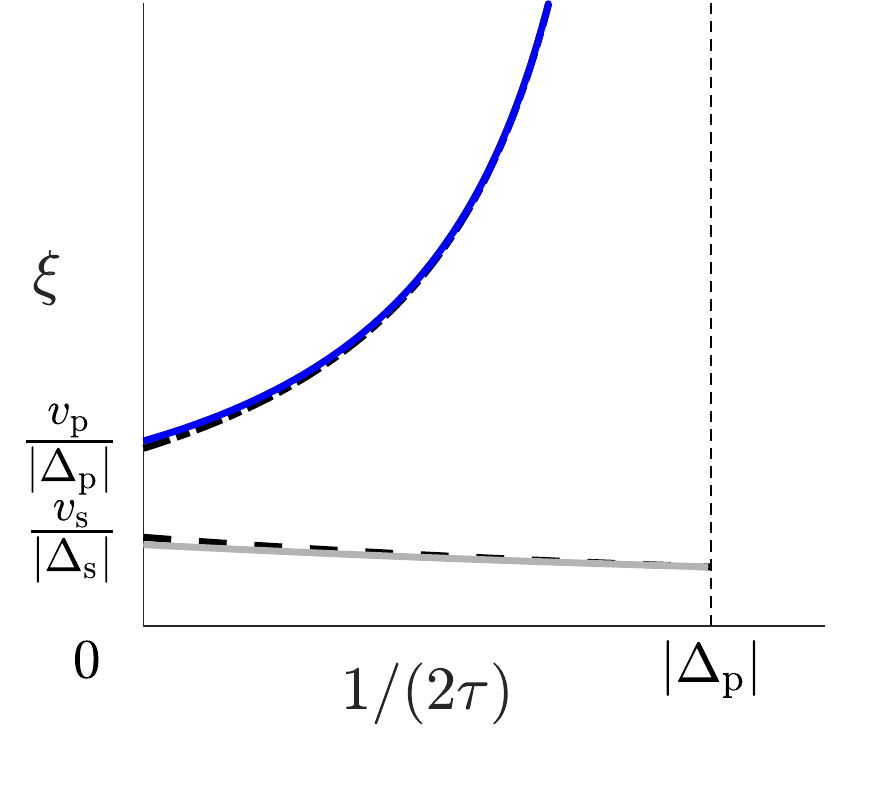}
		\llap{\hskip 2mm \parbox[c]{8cm}{\vspace{-6.7cm}(d)}}
		\tabularnewline
	\end{tabular}
	\vskip -5mm
	\caption{(a) Example of the low-energy spectrum of the normal strip 
		($|y|<W/2$), before it is coupled to the SCs.
		Red arrows show the spin expectation value, averaged over $y$,  $\int{\rm d}y \langle\boldsymbol{\sigma}(y)\rangle$. The direction of the arrow indicates the direction of the spin in the ($xy$) plane.
		(b) Upon introducing the superconductors, a gap is induced.
		For the parameters considered here,
		the junction supports five gapped transverse channels. 
		(c-d) Localization
		length versus disorder in a \emph{p}-wave SC, $\xi_{{\rm p}}$
		(black dotted line), and in an \emph{s}-wave SC, $\xi_{{\rm s}}$ (black
		dashed line). Disorder increases $\xi_{{\rm p}}$ and decreases $\xi_{{\rm s}}$.
		A multi-channel TSC can sometimes be viewed as a combination of a \emph{p}-wave
		SC and an \emph{s}-wave SC. The overall localization length is determined
		by the larger between $\xi_{{\rm p}}$ and $\xi_{{\rm s}}.$ 
		(c) When the
		\emph{p}-wave gap, $|\Delta_{{\rm p}}|$, exceeds the \emph{s}-wave
		gap, $|\Delta_{{\rm s}}|,$ the overall $\xi$ shows non-monotonic behavior as a function of disorder strength (blue solid line). This is the situation in the planar JJ studied
		here. (d) When $|\Delta_{{\rm p}}|\le|\Delta_{{\rm s}}|$, disorder causes $\xi$ to increase monotonically, reaching a phase transition at $1/2\tau_{{\rm c}}=|\Delta_{\rm p}|$.
		\label{fig:xi_s_wave_plus_p_wave}}
\end{figure}

We find that the enhancement of localization by weak disorder in the planar JJ is related to the structure of the low-energy excitations confined to the junction. The excitations carry a longitudinal momentum $k_x$. The spectrum is gapped, and the smallest gap is at large $k_x$, close the Fermi momenta of the 2DEG~\cite{Pientka2017topological}.  At these $k_x$'s spin-orbit
coupling (SOC) dominates over the Zeeman field, causing the spins of opposite-momenta
modes in each channel to be oppositely polarized, thereby suppressing the detrimental intra-channel backscattering~\cite{Brouwer2011topological}. Consequently, disorder effectively increases the gap of the large-momentum
channels. In contrast, at small $k_x$ Zeeman field
dominates over SOC, allowing for intra-channel backscattering, which
decreases the effective gap. The smallest of the gaps determines $\xi$. Weak disorder then increases the large momentum gap and enhances localization. As disorder is increased, the trend changes when the gaps at small and large momentum become equal  (see also Fig.~\ref{fig:xi_s_wave_plus_p_wave}).

We begin with a numerical analysis of the dependence of $\xi$ on disorder in a
planar JJ.
We then consider a low-energy model of a multi-channel
TSC. Finally, we construct a simplified model of the planar JJ which qualitatively reproduces the numerical results.

\paragraph*{Numerical analysis of the planar Josephson junction.\textemdash{}}

The planar JJ consists of two conventional
superconductors in proximity to a Rashba-spin-orbit-coupled 2DEG~\cite{Hart2017controlled}. The superconductors
are separated by a distance $W$, and are of length $L_{x}$ in the
$x$ direction, see Fig.~\hyperref[fig:PJJ]{\ref{fig:PJJ}}. As shown
theoretically~\cite{Hell2017two,Pientka2017topological}, by applying an in-plane magnetic field and
controlling the phase bias, the junction can realize
a 1D TSC. Experimental evidence for a TSC has been recently reported \cite{Ren2018topological,Fornieri2018evidence}.

In the presence of impurity-potential disorder, the system's Hamiltonian is
\begin{equation}
\begin{split}\mathcal{H} & =\left[-\frac{\nabla^{2}}{2m_{{\rm e}}}-\mu(y)+U(x,y)-i\alpha\left(\sigma_{y}\partial_{x}-\sigma_{x}\partial_{y}\right)\right]\tau_{z}\\
& +E_{{\rm Z}}(y)\sigma_{x}+{\rm Re[}\Delta(y)]\tau_{x}-{\rm Im}[\Delta(y)]\tau_{y},
\end{split}
\label{eq:H_PJJ}
\end{equation}
where $m_{{\rm e}}$ is the effective electron mass in the 2DEG, $\mu(y)=\mu_{{\rm J}}\theta(w/2-|y|)+\mu_{{\rm SC}}\theta(|y|-w/2)$
is the chemical potential, with $\mu_{{\rm J}}$ ($\mu_{{\rm SC}}$)
its value in the junction (below the superconductors), $\alpha$
is the Rashba spin-orbit coupling coefficient, $E_{{\rm Z}}(y)=E_{{\rm Z,J}}\theta(w/2-|y|)+E_{{\rm Z,SC}}\theta(|y|-w/2)$
is the Zeeman splitting due to the in-plane magnetic field, with $E_{{\rm Z,J}}$
($E_{{\rm Z,SC}}$) being its values in the junction (below the superconductors),
and $\Delta(y)=\Delta_{0}\theta(|y|-w/2)\exp[i{\rm sgn}(y)\phi/2]$
is the electrons' pairing potential, $\phi$ being the phase difference between the two superconductors.
Here, $U(x,y)$ is a random disorder potential having zero average
and short-range correlations, $\langle U(\boldsymbol{r})U(\boldsymbol{r}')\rangle=\delta(\boldsymbol{r}-\boldsymbol{r}')/(m_{\rm e}\tau)$, where $\tau$ is the mean free time for disorder scattering in the bare 2DEG. In writing Eq.~\eqref{fig:PJJ} we have used the
Nambu basis, $\Psi^{\dagger}(\boldsymbol{r})=[\psi_{\uparrow}^{\dagger}(\boldsymbol{r}),\psi_{\downarrow}^{\dagger}(\boldsymbol{r}),\psi_{\downarrow}(\boldsymbol{r}),-\psi_{\uparrow}(\boldsymbol{r})]$,
where $\psi_{s}^{\dagger}(\boldsymbol{r})$ creates an electron in
the 2DEG with spin $s$ at position $\boldsymbol{r}=(x,y)$. Accordingly,
the sets of Pauli matrices, $\sigma_{\alpha=x,y,z}$ and $\tau_{\alpha=x,y,z}$,
operate on the spin and particle-hole degrees of freedom, respectively.

To analyze the disordered system numerically, we
use a lattice model and construct a corresponding tight-binding
Hamiltonian. The topological invariant $\mathcal{Q}$ and the localization length $\xi$
can be obtained from the scattering matrix between two fictitious
leads at $x=0$ and $x=L_{x}$ (which extend throughout
the $y$ direction). The scattering matrix is calculated numerically using a recursive Green-function method~\cite{Lee1981Anderson,SM}.

Let $r(\varepsilon)$ be the reflection matrix for electrons and holes incident on the left at energy $\varepsilon$. The topological invariant satisfies~\cite{Akhmerov2011quantized,Fulga2011scattering}, $\mathcal{Q}={\rm det}[r(\varepsilon=0)]$, which
in the limit $L_{x}\to\infty$ takes the values $1$ in the trivial
phase and $-1$ in the topological phase.

We obtain $\xi$ from finite-size scaling of the zero-energy
transmission probability matrix, $T(0)=\mathbbm{1}-r^\dagger(0) r(0)$.
Except for the phase transition, the eigenvalues of $T(0)$ decay
exponentially with $L_{x}$~\cite{Beenakker1997Random,Evers2008Anderson}. The smallest exponent determines the localization length of mid-gap zero energy states.
In the topological phase, this defines the Majorana localization
length $\xi$. We average $\xi$ over many disorder realizations.

Figure~\hyperref[fig:PJJ]{\ref{fig:PJJ}(a)} presents the phase diagram of the clean system [$U(x,y)=0$], previously obtained in
Ref.~\cite{Pientka2017topological}. We note the chemical potential
need not be fine-tuned for the system to be topological; in particular, it can be substantially
larger than $E_{{\rm Z,J}}$. In the topological phase, the junction
hosts zero-energy Majorana bound states (MBS) at the junction's ends near $x=0$ and $x=L_x$.

Figure~\hyperref[fig:PJJ]{\ref{fig:PJJ}(b)} presents $\xi$ versus
disorder strength, represented by the inverse mean free time of the
underlying 2DEG, $\tau^{-1}$, for different values of $E_{\rm Z,J}$ and $\phi$ [see markers in Fig.~\hyperref[fig:PJJ]{\ref{fig:PJJ}(a)}]. In all cases shown, $\xi$ first decreases as a function of $\tau^{-1}$, reaching a minimum
which can be an order of magnitude smaller than its value in
the clean system. This makes the Majorana bound states more protected against perturbations
that can potentially couple them. When increasing disorder strength
further, $\xi$ eventually increases, diverging
at the phase transition to the trivial phase, shown by the vertical
dashed lines. Notice the phase
transition occurs at a critical disorder strength, $\tau_{{\rm c}}^{-1}$, much larger than $E_{{\rm gap}}$.

\paragraph*{\label{sec:Low_E_model}Low-energy model.\textemdash{}}

To understand the above results,
we consider a more general model of a 1D multi-channel TSC, comprising of linearly-dispersing electronic modes,
$\phi_{m}(x),$ and given by $H=H_{0}+H_{{\rm dis}}$, with
\begin{equation}
\begin{split} & H_{0}=\sum_{m=\pm1}^{\pm N}\int{\rm d}x\Big\{ -iv_{m}\phi_{m}^{\dagger}(x)\partial_{x}\phi_{m}(x)\\
& \hskip 23mm +\frac{1}{2}\left[\Delta_{m}\phi_{m}^{\dagger}(x)\phi_{-m}^{\dagger}(x)+{\rm h.c.}\right]\Big\},\\
& H_{{\rm dis}}=\sum_{m,n}\int{\rm d}xe^{i(k_{{\rm F},m}-k_{{\rm F},n})x}V_{mn}(x)\phi_{m}^{\dagger}(x)\phi_{n}(x).
\end{split}
\label{eq:H}
\end{equation}
Here each of the $N$ conducting channels contains a right-moving
mode ($v_{m>0}>0$) and a left-moving mode ($v_{m<0}<0$),
$k_{{\rm F},m}$ is the Fermi momentum of the $m$-th mode, $v_{m}$
is the mode velocity, $\Delta_{m}$ is a pairing potential in the $m$-th channel, and $V_{mn}(x)$ are scattering terms arising from disorder. Notice $V_{nm}^{\ast}(x)=V_{mn}(x)$ due to hermiticity, and $\Delta_{m}=-\Delta_{-m}$ due to the anticommutativity of $\{\phi_{m}\}$. In the clean limit, the system is topological for odd $N$, and trivial for even $N$. The Majorana localization length (for odd $N$) is determined by the maximal $\xi_{m}^{0}=v_{m}/|\Delta_{m}|.$

This model can be related to the planar JJ, at low energies, by first solving the Hamiltonian of Eq.~(\ref{eq:H_PJJ}) inside the junction ($|y|\le W/2$) in the absence of coupling to
the SCs, i.e., when the reflection from the SCs is purely normal [see e.g. Fig.~\hyperref[fig:xi_s_wave_plus_p_wave]{\ref{fig:xi_s_wave_plus_p_wave}(a)}], then linearizing the spectrum near the Fermi points to obtain
$\phi_{m}$ and $v_{m}$, and finally considering the induced superconductivity
in the form of the pairing potentials, $\Delta_{m}$~\cite{SM}. This is justified when the Fermi level is far enough from the bottom
of the band, compared with $|\Delta_{m}|$, $|V_{mn}|$. Omitting inter-channel pairings is justified whenever the energy mismatch, $\min(v_n,v_m)|k_{n}-k_{m}|,$ is large compared with
the inter-channel pairing.

In the above model, Eq. (\ref{eq:H}), we assume that $k_{{\rm F},-m}=-k_{{\rm F,}m}$, $v_{-m}=-v_{m}$, and  $V_{mn}=V_{-m,-n}$. This will indeed be the case in the planar JJ
due to a reflection symmetry, $\sigma_{x}\mathcal{H}(-x,y)\sigma_{x}=\mathcal{H}(x,y)$, present  \emph{in the clean limit}~\cite{SM}. The elements of the disorder matrix are normally distributed, with zero mean and short-range correlations, $\overline{V_{mn}(x)V_{mn}(x')} = \gamma_{mn}\delta(x-x')$, where the upper bar denotes disorder averaging, and $\gamma_{mn}$ is related to the disorder-induced transition rate from mode $m$ to $n$, through $\tau^{-1}_{mn}=|\gamma_{mn}/v_n|$. While $\gamma_{mn}$ is generally complex, in our case it may be chosen real and positive, thanks to a time-reversal-like symmetry, $\mathcal{H}^\ast(x,-y)=\mathcal{H}(x,y)$, which exists in the clean limit~\cite{Pientka2017topological,SM}. We make this choice here.

To obtain a correction to $\Delta_{m},$
in the form of a disorder self energy, we examine the
Nambu-Gor'kov Green function, $G_{mn}(x-x';i\omega)=\int{\rm d}\tau e^{-i\tau\omega}\overline{\langle\mathcal{T}_{\tau}\Phi_{m}(x',0)\Phi_{n}^{\dagger}(x,\tau)\rangle},$
where $\Phi_{m}^{\dagger}=(\phi_{m}^{\dagger},\phi_{-m})$. In the absence of disorder,
the momentum-space Green function reads
\begin{equation}
G_{mn}^{0}(q,i\omega=0)=\frac{-\delta_{mn}}{(v_{m}q)^{2}+|\Delta_{m}|^{2}}\begin{pmatrix}v_{m}q & \Delta_{m}\\
\Delta_{m}^{\ast} & -v_{m}q
\end{pmatrix}.
\end{equation}
For weak disorder, we can obtain the self-energy within the Born approximation~\cite{SM},
\begin{equation}
\begin{split}\Sigma_{m}(q,0)= & \sum_{n\neq m}|\gamma_{mn}|\int\frac{{\rm d}p}{2\pi}e^{i\frac{\alpha_{mn}}{2}\tau_z}\tau_{z}G_{nn}^{0}(p,0)\tau_{z}e^{-i\frac{\alpha_{mn}}{2}\tau_z}\\
= & \sum_{n\neq m}\frac{1}{2\tau_{mn}}e^{i[\arg(\Delta_n)+\alpha_{mn}]\tau_z}\tau_x,
\end{split}
\label{eq:self_energy}
\end{equation}
where $\alpha_{mn}\equiv\arg(\gamma_{mn})$. Comparing with the unperturbed Green function, we see that disorder
changes the effective pairing potentials according to
\begin{equation}
\Delta_{m}^{{\rm eff}}=\Delta_{m}+\frac{1}{2}\sum_{n\neq m}\frac{1}{\tau_{mn}}e^{i[\arg(\Delta_{n})+\alpha_{mn}]}.\label{eq:Delta_m_eff}
\end{equation}
Notice that the contribution of mode $n$ to $|\Delta_{m}|$ depends
on the inter-channel scattering rate, $\tau^{-1}_{mn}$, the scattering phase, $\alpha_{mn}$, {\it and} on the relative phase between $\Delta_{m}$ and $\Delta_{n}$.
Importantly, disorder can either decrease or increase $|\Delta_{m}|$ (and therefore
increase or decrease $\xi_{m}$). The process underlying Eq. (\ref{eq:Delta_m_eff})
is depicted in Fig.~\hyperref[fig:Low-energy-models]{\ref{fig:Low-energy-models}}.

\paragraph*{Disordered s-wave vs. disordered p-wave superconductor.\textemdash{}}

We explore
two special cases of the multi-channel superconductor: (i) a single-channel
\emph{p}-wave SC and (ii) a single-(spinful)-channel \emph{s}-wave
SC. These cases clarify the non-monotonic behavior of $\xi$ for the disordered planar JJ,
observed in Fig.~\hyperref[fig:PJJ]{\ref{fig:PJJ}(b)}.

The low-energy Hamiltonian of a single-channel \emph{p}-wave SC is
obtained by setting $N=1$ in Eq.~\eqref{eq:H}, with $v_1\equiv v_{\rm p}$, $\Delta_{1}=-\Delta_{-1}\equiv \Delta_{\rm p}$, and $\tau_{1,-1}\equiv \tau_{\rm p}$.
Equation~(\ref{eq:Delta_m_eff}) then yields
\begin{equation}
|\Delta_{{\rm p}}^{{\rm eff}}|=|\Delta_{{\rm p}}|-1/2\tau_{\rm p}.
\label{eq:delta_p_eff}
\end{equation}
The localization length can then be obtained by $\xi_{{\rm p}}=v_{\rm p}/|\Delta_{{\rm p}}^{{\rm eff}}|$, yielding the known result~\cite{Brouwer2011Probability}, $1/\xi_{{\rm p}}=1/\xi_{{\rm p}}^{0}-1/2l_{\rm p},\label{eq:p_wave_xi}$ where $l_{\rm p}=v_{\rm p}\tau_{\rm p}$ is the mean free path, and $\xi_{{\rm p}}^{0}=v_{\rm p}/|\Delta_{{\rm p}}|$.

For a single-channel $s$-wave superconductor there are no zero-energy
end modes and $\xi$ is the length to which a single electron at zero
energy penetrates the superconductor before being reflected. The index
$m=1,2$ corresponds to the two spin directions. The spin-singlet nature of the pairing dictates $\Delta_{1}=-\Delta_{2}=\Delta_{\rm s}$
and the spin-independence of the disorder forbids intra-channel scattering
and dictates $\tau^{-1}_{1,-1}=\tau^{-1}_{2,-2}=0$. Furthermore, the
two velocities are the same, $v_{1}=v_{2}=v_{\rm s}.$

Setting this in Eq. (\ref{eq:Delta_m_eff}), we have
\begin{equation}
|\Delta_{{\rm s}}^{{\rm eff}}|=|\Delta_{{\rm s}}|+1/2\tau^{-1}_{\rm s},\label{eq:delta_s_eff}
\end{equation}
where $\tau_{11}=\tau_{22}\equiv\tau_{\rm s}$, and correspondingly $1/\xi_{{\rm s}}=1/\xi_{{\rm s}}^{0}+1/2l_{\rm s}$, where $l_{\rm s}=v_{\rm s}\tau_{\rm s}$. Unlike the case of the single-channel \emph{p}-wave SC, the localization
length in the \emph{s}-wave case \emph{decreases} monotonically. We emphasize
that the relative sign difference in Eq. (\ref{eq:delta_p_eff}), compared
to Eq. (\ref{eq:delta_s_eff}), stems from (i) lack of scattering between opposite spins and (ii) the \emph{s}-wave spin-singlet nature of the pairing. While these results for $\xi_{{\rm p}}$ and $\xi_{{\rm s}}$ were obtained using
a weak-disorder perturbative analysis, they are actually exact for
the linearized model of Eq. (\ref{fig:Low-energy-models}), as shown in the Supplemental Material~\cite{SM}.

The results for $\xi_{{\rm p}}$ and $\xi_{{\rm s}}$ let us understand the non-monotonic behavior of $\xi$
in the planar JJ [see Fig.~\hyperref[fig:PJJ]{\ref{fig:PJJ}(b)}]. The low-energy spectrum of sub-gap excitations confined between the two superconductors may be seen as coming out of superconducting pairing of several low-energy modes. Figure~\hyperref[fig:xi_s_wave_plus_p_wave]{\ref{fig:xi_s_wave_plus_p_wave}(a)} presents an example of the spectrum of the Hamiltonian, Eq. (\ref{eq:H_PJJ}), confined within the junction under the assumption of full normal reflection. The red arrows, representing the spin expectation
values of the modes, indicate that the outer channels (larger
Fermi momentum) are largely spin polarized due to the spin-orbit coupling, with opposite-momentum modes
having approximately opposite spins. For the inner channel, the spin varies along the $y$--direction, resulting in a smaller expectation value.

Since pairing is induced by an \emph{s}-wave SC, the large-momentum channels, being spin-polarized,
behave as a \emph{s}-wave SC, with a localization length,
$\xi_{{\rm s}}$, that decreases with disorder. In contrast, the small-momentum channel is not spin-polarized,  and  allows
for intra-channel backscattering. Consequently this channel behaves as a
disordered \emph{p}-wave SC, with localization length, $\xi_{{\rm p}}$, that increases with disorder. The overall localization length of the system,
$\xi$, is then the larger between $\xi_{{\rm s}}$
and $\xi_{{\rm p}}$.

The behavior of $\xi$ versus disorder therefore depends on the relative size of $|\Delta_{{\rm p}}|$ and $|\Delta_{{\rm s}}|$. Assuming, for simplicity $\tau_{\rm p}\sim\tau_{\rm s}\sim\tau$, and $v_{\rm s}\sim v_{\rm p}$, we find that when $|\Delta_{\rm p}|>|\Delta_{\rm s}|$ [Fig.~\hyperref[fig:xi_s_wave_plus_p_wave]{\ref{fig:xi_s_wave_plus_p_wave}(c)}], the localization length {\it decreases} for weak disorder.
With stronger disorder  the two gaps approach one another. Consequently,
scattering between the large-momentum
and the low-momentum channels causes ``level repulsion'' between
$\xi_{{\rm s}}$ and $\xi_{{\rm p}}$, as depicted in
Fig.~\hyperref[fig:xi_s_wave_plus_p_wave]{\ref{fig:xi_s_wave_plus_p_wave}(c,d)}, and $\xi$ increases with disorder (blue solid line). Notice that the critical
disorder strength, $1/2\tau_{{\rm c}}=|\Delta_{{\rm p}}|$, can be
much larger than the gap of the clean system, $|\Delta_{{\rm s}}$|. In contrast, if $|\Delta_{{\rm p}}|\le|\Delta_{{\rm s}}|$ [Fig.~\hyperref[fig:xi_s_wave_plus_p_wave]{\ref{fig:xi_s_wave_plus_p_wave}(d)}], disorder causes $\xi$ to increase monotonically,
diverging at the critical disorder $1/2\tau_{{\rm c}}=|\Delta_{\rm p}|$,
which now equals the gap of the clean system.

In the planar JJ, the gap of the large-momentum channels
is indeed the smaller one [see Fig.~\hyperref[fig:xi_s_wave_plus_p_wave]{\ref{fig:xi_s_wave_plus_p_wave}(b)}]. For a not-too-narrow junction, the low-momentum gap is approximately $\min(\Delta_0,hv_F/W)$, while the large-momentum gap is approximately $\hbar^2/2m_eW^2$.
This difference may be viewed as originating from the fact that high-momentum electrons propagate almost parallel to the SCs, and are therefore only weakly coupled to the SCs. For this system, then, disorder may increase the effective gap from the scale of $\hbar^2/2m_eW^2$ to the scale of $\Delta_0$.

\begin{figure}
	\begin{tabular}{lr}
		\includegraphics[clip=true,trim =10mm 0mm 2mm 0mm,scale=0.235]{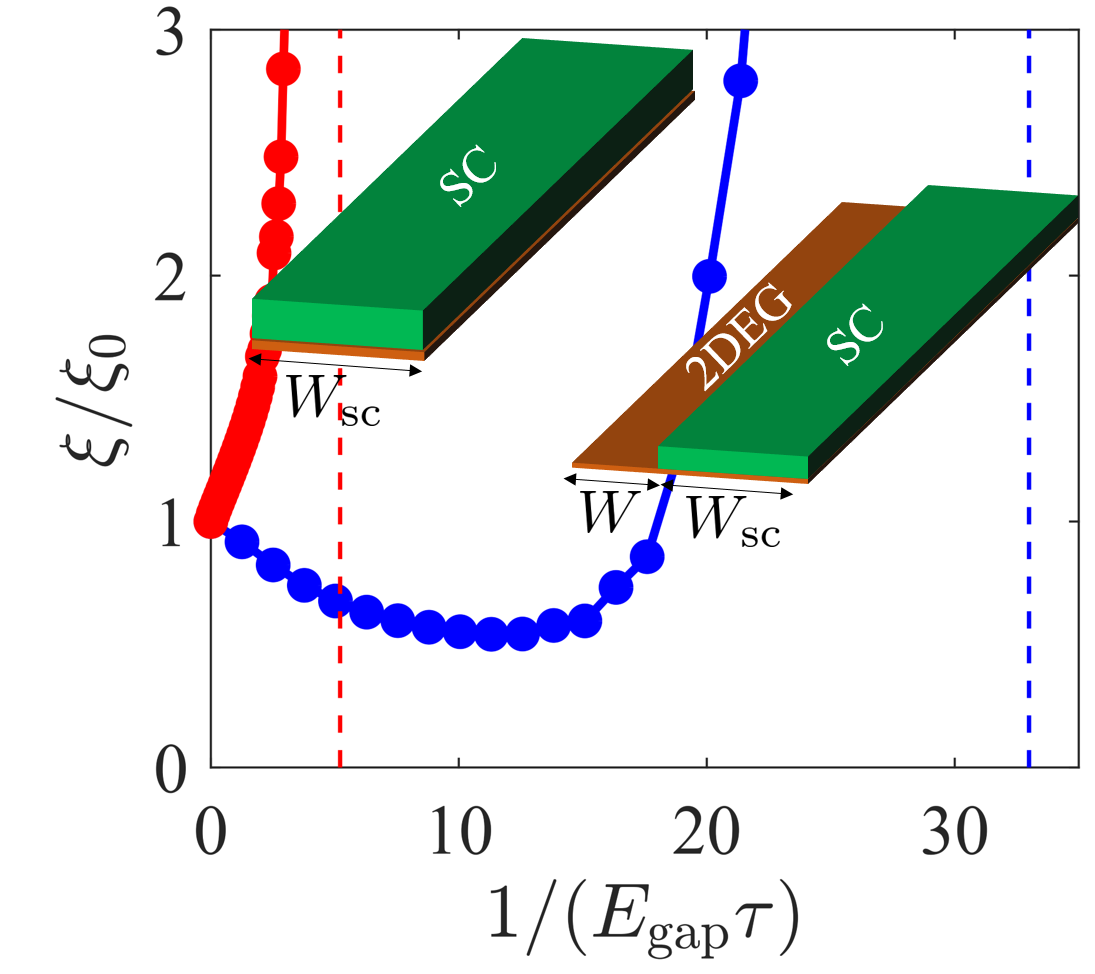}
		\llap{\hskip 2mm \parbox[c]{8.25cm}{\vspace{-7cm}(a)}}
		&
		\hskip -1mm
		\includegraphics[clip=true,trim =0mm 0mm 0mm 0mm,scale=0.485]{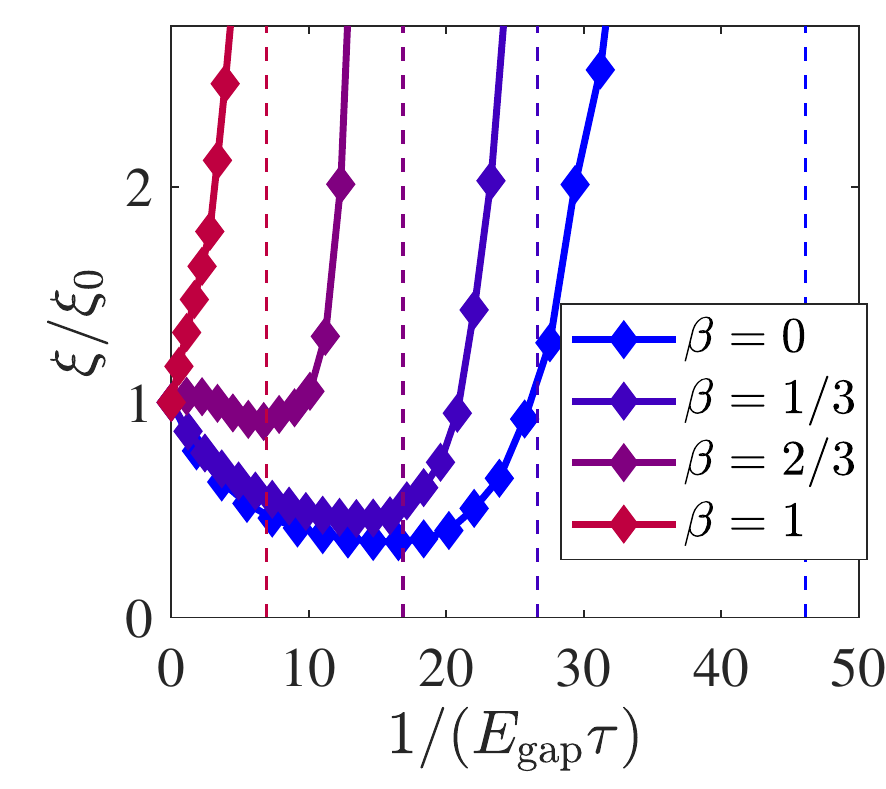}
		\llap{\hskip 2mm \parbox[c]{8.1cm}{\vspace{-7cm}(b)}}
		\tabularnewline
	\end{tabular}
	\vskip -3mm
	\caption{(a) Majorana localization length versus disorder in a 2DEG
		coupled to an \emph{s}-wave SC in two different geometries. In the first
		geometry (blue), the smaller superconducting gap is at the large-momentum
		channels, while in the second geometry (red) the gaps of the different channels are of the same order. The system's parameters are as in Fig.~\ref{fig:PJJ}, with $E_{\rm Z,J}=1$, and with $\Delta_0=1$ ($\Delta_0=0.25$) for the first (second) geometry (The reduced $\Delta_0$ in the second geometry is necessary for the system to be topological). (b) Majorana localization length for the planar JJ
		{[}see Eq. (\ref{eq:H_PJJ}) and Fig. \ref{fig:PJJ}] for different
		ratios of magnetic-disorder strength to potential-disorder
		strength. Magnetic disorder couples opposite-spin modes. The rest of the system's parameters are as in Fig.~\ref{fig:PJJ}, with $\phi=\pi$, $E_{\rm Z,J}=1$.
		\label{fig:mag_imp}}
\end{figure}

We test our understanding by studying two geometries of superconducting
proximity [Fig.~\hyperref[fig:mag_imp]{\ref{fig:mag_imp}(a)}], where a strip of 2DEG is coupled to a single
SC from the side (blue line) or from above the strip (red line). While in the former
the large-momentum (\emph{s}-wave) gap is the smallest, giving
rise to behavior similar to the planar JJ, this is not the case in the latter geometry, resulting in a monotonically-increasing $\xi$.

As another test, we add to Eq.~\eqref{eq:H_PJJ} a magnetic disorder term $\mathcal{H}_{{\rm m}}=U_{{\rm m}}(\boldsymbol{r})\sigma_{z}$. Here  $U_{m}(\boldsymbol{r})$ is a random
field with zero average and correlations $\langle U_{{\rm m}}(\boldsymbol{r})U_{{\rm m}}(\boldsymbol{r}')\rangle=\gamma_{{\rm m}}\theta(W/2-|y|)\delta(\boldsymbol{r}-\boldsymbol{r}')$~\footnote{We limit $U_{m}(\boldsymbol{r})$ to the junction since we are not interested here in its effect on the SCs.}. 
Figure~\hyperref[fig:mag_imp]{\ref{fig:mag_imp}(b)} presents $\xi$ for different values of the ratio between magnetic and potential
disorder, $\beta=\gamma_{{\rm m}}/\gamma$, where $\gamma=1/(m_{\rm e}\tau)$. Since magnetic disorder
can scatter between the opposite-spin states, the large-momentum channels
do not behave anymore as an \emph{s}-wave SC, and instead are more
similar to a multi-channel \emph{p}-wave SC~\cite{Rieder2013reentrant,Rieder2014Density,Potter2010multichannel}. Indeed, with increasing $\beta$, the disorder-induced decrease in $\xi$ diminishes.

\paragraph*{Acknowledgments.\textemdash{}}
We have benefited from the insightful comments of B Halperin and Y
Oreg. We also thank M Buchhold for useful discussions. We acknowledge
support from the Walter Burke Institute for Theoretical Physics at
Caltech (AH), the Israel Science Foundation (AS), the European Research Council under the European
Community Seventh Framework Program (FP7/2007-
2013)/ERC Project MUNATOP (AS), Microsoft Station
Q (AS), and the DFG (CRC/Transregio 183, EI 519/7-1) (AS).

\bibliographystyle{apsrev4-1}
\bibliography{References_DESD_ver_2}

\begin{thebibliography}{41}%
\makeatletter
\providecommand \@ifxundefined [1]{%
 \@ifx{#1\undefined}
}%
\providecommand \@ifnum [1]{%
 \ifnum #1\expandafter \@firstoftwo
 \else \expandafter \@secondoftwo
 \fi
}%
\providecommand \@ifx [1]{%
 \ifx #1\expandafter \@firstoftwo
 \else \expandafter \@secondoftwo
 \fi
}%
\providecommand \natexlab [1]{#1}%
\providecommand \enquote  [1]{``#1''}%
\providecommand \bibnamefont  [1]{#1}%
\providecommand \bibfnamefont [1]{#1}%
\providecommand \citenamefont [1]{#1}%
\providecommand \href@noop [0]{\@secondoftwo}%
\providecommand \href [0]{\begingroup \@sanitize@url \@href}%
\providecommand \@href[1]{\@@startlink{#1}\@@href}%
\providecommand \@@href[1]{\endgroup#1\@@endlink}%
\providecommand \@sanitize@url [0]{\catcode `\\12\catcode `\$12\catcode
  `\&12\catcode `\#12\catcode `\^12\catcode `\_12\catcode `\%12\relax}%
\providecommand \@@startlink[1]{}%
\providecommand \@@endlink[0]{}%
\providecommand \url  [0]{\begingroup\@sanitize@url \@url }%
\providecommand \@url [1]{\endgroup\@href {#1}{\urlprefix }}%
\providecommand \urlprefix  [0]{URL }%
\providecommand \Eprint [0]{\href }%
\providecommand \doibase [0]{http://dx.doi.org/}%
\providecommand \selectlanguage [0]{\@gobble}%
\providecommand \bibinfo  [0]{\@secondoftwo}%
\providecommand \bibfield  [0]{\@secondoftwo}%
\providecommand \translation [1]{[#1]}%
\providecommand \BibitemOpen [0]{}%
\providecommand \bibitemStop [0]{}%
\providecommand \bibitemNoStop [0]{.\EOS\space}%
\providecommand \EOS [0]{\spacefactor3000\relax}%
\providecommand \BibitemShut  [1]{\csname bibitem#1\endcsname}%
\let\auto@bib@innerbib\@empty
\bibitem [{\citenamefont {Kitaev}(2001)}]{Kitaev2001unpaired}%
  \BibitemOpen
  \bibfield  {author} {\bibinfo {author} {\bibfnamefont {A.}~\bibnamefont
  {Kitaev}},\ }\href
  {http://iopscience.iop.org/article/10.1070/1063-7869/44/10S/S29/meta}
  {\bibfield  {journal} {\bibinfo  {journal} {Phys. Usp.}\ }\textbf {\bibinfo
  {volume} {44}},\ \bibinfo {pages} {131} (\bibinfo {year} {2001})}\BibitemShut
  {NoStop}%
\bibitem [{\citenamefont {Read}\ and\ \citenamefont
  {Green}(2000)}]{Read2000paired}%
  \BibitemOpen
  \bibfield  {author} {\bibinfo {author} {\bibfnamefont {N.}~\bibnamefont
  {Read}}\ and\ \bibinfo {author} {\bibfnamefont {D.}~\bibnamefont {Green}},\
  }\href {http://link.aps.org/doi/10.1103/PhysRevB.61.10267} {\bibfield
  {journal} {\bibinfo  {journal} {Phys. Rev. B}\ }\textbf {\bibinfo {volume}
  {61}},\ \bibinfo {pages} {10267} (\bibinfo {year} {2000})}\BibitemShut
  {NoStop}%
\bibitem [{\citenamefont {Qi}\ and\ \citenamefont
  {Zhang}(2011)}]{Qi2011topological}%
  \BibitemOpen
  \bibfield  {author} {\bibinfo {author} {\bibfnamefont {X.-L.}\ \bibnamefont
  {Qi}}\ and\ \bibinfo {author} {\bibfnamefont {S.-C.}\ \bibnamefont {Zhang}},\
  }\href {\doibase 10.1103/RevModPhys.83.1057} {\bibfield  {journal} {\bibinfo
  {journal} {Rev. Mod. Phys.}\ }\textbf {\bibinfo {volume} {83}},\ \bibinfo
  {pages} {1057} (\bibinfo {year} {2011})}\BibitemShut {NoStop}%
\bibitem [{\citenamefont {Alicea}(2012)}]{Alicea2012}%
  \BibitemOpen
  \bibfield  {author} {\bibinfo {author} {\bibfnamefont {J.}~\bibnamefont
  {Alicea}},\ }\href {\doibase 10.1088/0034-4885/75/7/076501} {\bibfield
  {journal} {\bibinfo  {journal} {Rep. Prog. Phys.}\ }\textbf {\bibinfo
  {volume} {75}},\ \bibinfo {pages} {076501} (\bibinfo {year}
  {2012})}\BibitemShut {NoStop}%
\bibitem [{\citenamefont {Beenakker}(2013)}]{Beenakker2013}%
  \BibitemOpen
  \bibfield  {author} {\bibinfo {author} {\bibfnamefont {C.~W.~J.}\
  \bibnamefont {Beenakker}},\ }\href {\doibase
  10.1146/annurev-conmatphys-030212-184337} {\bibfield  {journal} {\bibinfo
  {journal} {Annu. Rev. Condens. Matt. Phys.}\ }\textbf {\bibinfo {volume}
  {4}},\ \bibinfo {pages} {113} (\bibinfo {year} {2013})}\BibitemShut {NoStop}%
\bibitem [{\citenamefont {Lutchyn}\ \emph {et~al.}(2018)\citenamefont
  {Lutchyn}, \citenamefont {Bakkers}, \citenamefont {Kouwenhoven},
  \citenamefont {Krogstrup}, \citenamefont {Marcus},\ and\ \citenamefont
  {Oreg}}]{Lutchyn2018majorana}%
  \BibitemOpen
  \bibfield  {author} {\bibinfo {author} {\bibfnamefont {R.}~\bibnamefont
  {Lutchyn}}, \bibinfo {author} {\bibfnamefont {E.}~\bibnamefont {Bakkers}},
  \bibinfo {author} {\bibfnamefont {L.}~\bibnamefont {Kouwenhoven}}, \bibinfo
  {author} {\bibfnamefont {P.}~\bibnamefont {Krogstrup}}, \bibinfo {author}
  {\bibfnamefont {C.}~\bibnamefont {Marcus}}, \ and\ \bibinfo {author}
  {\bibfnamefont {Y.}~\bibnamefont {Oreg}},\ }\href
  {https://www.nature.com/articles/s41578-018-0003-1} {\bibfield  {journal}
  {\bibinfo  {journal} {Nat. Rev. Mat.}\ }\textbf {\bibinfo {volume} {3}},\
  \bibinfo {pages} {52} (\bibinfo {year} {2018})}\BibitemShut {NoStop}%
\bibitem [{\citenamefont {Aguado}(2017)}]{Aguado2017majorana}%
  \BibitemOpen
  \bibfield  {author} {\bibinfo {author} {\bibfnamefont {R.}~\bibnamefont
  {Aguado}},\ }\href
  {https://www.sif.it/riviste/sif/ncr/econtents/2017/040/11/article/0}
  {\bibfield  {journal} {\bibinfo  {journal} {La Rivista Del Nuovo Cimento}\
  }\textbf {\bibinfo {volume} {40}},\ \bibinfo {pages} {523} (\bibinfo {year}
  {2017})}\BibitemShut {NoStop}%
\bibitem [{\citenamefont {Motrunich}\ \emph {et~al.}(2001)\citenamefont
  {Motrunich}, \citenamefont {Damle},\ and\ \citenamefont
  {Huse}}]{Motrunich2001Griffiths}%
  \BibitemOpen
  \bibfield  {author} {\bibinfo {author} {\bibfnamefont {O.}~\bibnamefont
  {Motrunich}}, \bibinfo {author} {\bibfnamefont {K.}~\bibnamefont {Damle}}, \
  and\ \bibinfo {author} {\bibfnamefont {D.~A.}\ \bibnamefont {Huse}},\ }\href
  {\doibase 10.1103/PhysRevB.63.224204} {\bibfield  {journal} {\bibinfo
  {journal} {Phys. Rev. B}\ }\textbf {\bibinfo {volume} {63}},\ \bibinfo
  {pages} {224204} (\bibinfo {year} {2001})}\BibitemShut {NoStop}%
\bibitem [{\citenamefont {Brouwer}\ \emph
  {et~al.}(2011{\natexlab{a}})\citenamefont {Brouwer}, \citenamefont
  {Duckheim}, \citenamefont {Romito},\ and\ \citenamefont {von
  Oppen}}]{Brouwer2011Probability}%
  \BibitemOpen
  \bibfield  {author} {\bibinfo {author} {\bibfnamefont {P.~W.}\ \bibnamefont
  {Brouwer}}, \bibinfo {author} {\bibfnamefont {M.}~\bibnamefont {Duckheim}},
  \bibinfo {author} {\bibfnamefont {A.}~\bibnamefont {Romito}}, \ and\ \bibinfo
  {author} {\bibfnamefont {F.}~\bibnamefont {von Oppen}},\ }\href {\doibase
  10.1103/PhysRevLett.107.196804} {\bibfield  {journal} {\bibinfo  {journal}
  {Phys. Rev. Lett.}\ }\textbf {\bibinfo {volume} {107}},\ \bibinfo {pages}
  {196804} (\bibinfo {year} {2011}{\natexlab{a}})}\BibitemShut {NoStop}%
\bibitem [{\citenamefont {Lobos}\ \emph {et~al.}(2012)\citenamefont {Lobos},
  \citenamefont {Lutchyn},\ and\ \citenamefont
  {Das~Sarma}}]{Lobos2012interplay}%
  \BibitemOpen
  \bibfield  {author} {\bibinfo {author} {\bibfnamefont {A.~M.}\ \bibnamefont
  {Lobos}}, \bibinfo {author} {\bibfnamefont {R.~M.}\ \bibnamefont {Lutchyn}},
  \ and\ \bibinfo {author} {\bibfnamefont {S.}~\bibnamefont {Das~Sarma}},\
  }\href {\doibase 10.1103/PhysRevLett.109.146403} {\bibfield  {journal}
  {\bibinfo  {journal} {Phys. Rev. Lett.}\ }\textbf {\bibinfo {volume} {109}},\
  \bibinfo {pages} {146403} (\bibinfo {year} {2012})}\BibitemShut {NoStop}%
\bibitem [{\citenamefont {Pientka}\ \emph {et~al.}(2013)\citenamefont
  {Pientka}, \citenamefont {Romito}, \citenamefont {Duckheim}, \citenamefont
  {Oreg},\ and\ \citenamefont {von Oppen}}]{Pientka2013signatures}%
  \BibitemOpen
  \bibfield  {author} {\bibinfo {author} {\bibfnamefont {F.}~\bibnamefont
  {Pientka}}, \bibinfo {author} {\bibfnamefont {A.}~\bibnamefont {Romito}},
  \bibinfo {author} {\bibfnamefont {M.}~\bibnamefont {Duckheim}}, \bibinfo
  {author} {\bibfnamefont {Y.}~\bibnamefont {Oreg}}, \ and\ \bibinfo {author}
  {\bibfnamefont {F.}~\bibnamefont {von Oppen}},\ }\href
  {http://stacks.iop.org/1367-2630/15/i=2/a=025001} {\bibfield  {journal}
  {\bibinfo  {journal} {New J. Phys.}\ }\textbf {\bibinfo {volume} {15}},\
  \bibinfo {pages} {025001} (\bibinfo {year} {2013})}\BibitemShut {NoStop}%
\bibitem [{\citenamefont {Huse}\ \emph {et~al.}(2013)\citenamefont {Huse},
  \citenamefont {Nandkishore}, \citenamefont {Oganesyan}, \citenamefont {Pal},\
  and\ \citenamefont {Sondhi}}]{Huse2013localization}%
  \BibitemOpen
  \bibfield  {author} {\bibinfo {author} {\bibfnamefont {D.~A.}\ \bibnamefont
  {Huse}}, \bibinfo {author} {\bibfnamefont {R.}~\bibnamefont {Nandkishore}},
  \bibinfo {author} {\bibfnamefont {V.}~\bibnamefont {Oganesyan}}, \bibinfo
  {author} {\bibfnamefont {A.}~\bibnamefont {Pal}}, \ and\ \bibinfo {author}
  {\bibfnamefont {S.~L.}\ \bibnamefont {Sondhi}},\ }\href {\doibase
  10.1103/PhysRevB.88.014206} {\bibfield  {journal} {\bibinfo  {journal} {Phys.
  Rev. B}\ }\textbf {\bibinfo {volume} {88}},\ \bibinfo {pages} {014206}
  (\bibinfo {year} {2013})}\BibitemShut {NoStop}%
\bibitem [{\citenamefont {Adagideli}\ \emph {et~al.}(2014)\citenamefont
  {Adagideli}, \citenamefont {Wimmer},\ and\ \citenamefont
  {Teker}}]{Adagideli2014effects}%
  \BibitemOpen
  \bibfield  {author} {\bibinfo {author} {\bibfnamefont {I.}~\bibnamefont
  {Adagideli}}, \bibinfo {author} {\bibfnamefont {M.}~\bibnamefont {Wimmer}}, \
  and\ \bibinfo {author} {\bibfnamefont {A.}~\bibnamefont {Teker}},\ }\href
  {\doibase 10.1103/PhysRevB.89.144506} {\bibfield  {journal} {\bibinfo
  {journal} {Phys. Rev. B}\ }\textbf {\bibinfo {volume} {89}},\ \bibinfo
  {pages} {144506} (\bibinfo {year} {2014})}\BibitemShut {NoStop}%
\bibitem [{Note1()}]{Note1}%
  \BibitemOpen
  \bibinfo {note} {This result is valid when the Fermi energy is large compared
  with $E_{\protect \rm gap}$, which is the limit of interest here. For the
  opposite limit see Ref.~\cite {Pientka2013signatures}.}\BibitemShut {Stop}%
\bibitem [{\citenamefont {Potter}\ and\ \citenamefont
  {Lee}(2010)}]{Potter2010multichannel}%
  \BibitemOpen
  \bibfield  {author} {\bibinfo {author} {\bibfnamefont {A.~C.}\ \bibnamefont
  {Potter}}\ and\ \bibinfo {author} {\bibfnamefont {P.~A.}\ \bibnamefont
  {Lee}},\ }\href {\doibase 10.1103/PhysRevLett.105.227003} {\bibfield
  {journal} {\bibinfo  {journal} {Phys. Rev. Lett.}\ }\textbf {\bibinfo
  {volume} {105}},\ \bibinfo {pages} {227003} (\bibinfo {year}
  {2010})}\BibitemShut {NoStop}%
\bibitem [{\citenamefont {Rieder}\ \emph {et~al.}(2013)\citenamefont {Rieder},
  \citenamefont {Brouwer},\ and\ \citenamefont
  {Adagideli}}]{Rieder2013reentrant}%
  \BibitemOpen
  \bibfield  {author} {\bibinfo {author} {\bibfnamefont {M.-T.}\ \bibnamefont
  {Rieder}}, \bibinfo {author} {\bibfnamefont {P.~W.}\ \bibnamefont {Brouwer}},
  \ and\ \bibinfo {author} {\bibfnamefont {I.}~\bibnamefont {Adagideli}},\
  }\href {\doibase 10.1103/PhysRevB.88.060509} {\bibfield  {journal} {\bibinfo
  {journal} {Phys. Rev. B}\ }\textbf {\bibinfo {volume} {88}},\ \bibinfo
  {pages} {060509} (\bibinfo {year} {2013})}\BibitemShut {NoStop}%
\bibitem [{\citenamefont {Rieder}\ and\ \citenamefont
  {Brouwer}(2014)}]{Rieder2014Density}%
  \BibitemOpen
  \bibfield  {author} {\bibinfo {author} {\bibfnamefont {M.-T.}\ \bibnamefont
  {Rieder}}\ and\ \bibinfo {author} {\bibfnamefont {P.~W.}\ \bibnamefont
  {Brouwer}},\ }\href {\doibase 10.1103/PhysRevB.90.205404} {\bibfield
  {journal} {\bibinfo  {journal} {Phys. Rev. B}\ }\textbf {\bibinfo {volume}
  {90}},\ \bibinfo {pages} {205404} (\bibinfo {year} {2014})}\BibitemShut
  {NoStop}%
\bibitem [{\citenamefont {Lu}\ \emph {et~al.}(2016)\citenamefont {Lu},
  \citenamefont {Burset}, \citenamefont {Tanuma}, \citenamefont {Golubov},
  \citenamefont {Asano},\ and\ \citenamefont {Tanaka}}]{Lu2016influence}%
  \BibitemOpen
  \bibfield  {author} {\bibinfo {author} {\bibfnamefont {B.}~\bibnamefont
  {Lu}}, \bibinfo {author} {\bibfnamefont {P.}~\bibnamefont {Burset}}, \bibinfo
  {author} {\bibfnamefont {Y.}~\bibnamefont {Tanuma}}, \bibinfo {author}
  {\bibfnamefont {A.~A.}\ \bibnamefont {Golubov}}, \bibinfo {author}
  {\bibfnamefont {Y.}~\bibnamefont {Asano}}, \ and\ \bibinfo {author}
  {\bibfnamefont {Y.}~\bibnamefont {Tanaka}},\ }\href {\doibase
  10.1103/PhysRevB.94.014504} {\bibfield  {journal} {\bibinfo  {journal} {Phys.
  Rev. B}\ }\textbf {\bibinfo {volume} {94}},\ \bibinfo {pages} {014504}
  (\bibinfo {year} {2016})}\BibitemShut {NoStop}%
\bibitem [{\citenamefont {Burset}\ \emph {et~al.}(2017)\citenamefont {Burset},
  \citenamefont {Lu}, \citenamefont {Tamura},\ and\ \citenamefont
  {Tanaka}}]{Burset2017current}%
  \BibitemOpen
  \bibfield  {author} {\bibinfo {author} {\bibfnamefont {P.}~\bibnamefont
  {Burset}}, \bibinfo {author} {\bibfnamefont {B.}~\bibnamefont {Lu}}, \bibinfo
  {author} {\bibfnamefont {S.}~\bibnamefont {Tamura}}, \ and\ \bibinfo {author}
  {\bibfnamefont {Y.}~\bibnamefont {Tanaka}},\ }\href {\doibase
  10.1103/PhysRevB.95.224502} {\bibfield  {journal} {\bibinfo  {journal} {Phys.
  Rev. B}\ }\textbf {\bibinfo {volume} {95}},\ \bibinfo {pages} {224502}
  (\bibinfo {year} {2017})}\BibitemShut {NoStop}%
\bibitem [{\citenamefont {Hell}\ \emph
  {et~al.}(2017{\natexlab{a}})\citenamefont {Hell}, \citenamefont {Leijnse},\
  and\ \citenamefont {Flensberg}}]{Hell2017two}%
  \BibitemOpen
  \bibfield  {author} {\bibinfo {author} {\bibfnamefont {M.}~\bibnamefont
  {Hell}}, \bibinfo {author} {\bibfnamefont {M.}~\bibnamefont {Leijnse}}, \
  and\ \bibinfo {author} {\bibfnamefont {K.}~\bibnamefont {Flensberg}},\ }\href
  {\doibase 10.1103/PhysRevLett.118.107701} {\bibfield  {journal} {\bibinfo
  {journal} {Phys. Rev. Lett.}\ }\textbf {\bibinfo {volume} {118}},\ \bibinfo
  {pages} {107701} (\bibinfo {year} {2017}{\natexlab{a}})}\BibitemShut
  {NoStop}%
\bibitem [{\citenamefont {Pientka}\ \emph {et~al.}(2017)\citenamefont
  {Pientka}, \citenamefont {Keselman}, \citenamefont {Berg}, \citenamefont
  {Yacoby}, \citenamefont {Stern},\ and\ \citenamefont
  {Halperin}}]{Pientka2017topological}%
  \BibitemOpen
  \bibfield  {author} {\bibinfo {author} {\bibfnamefont {F.}~\bibnamefont
  {Pientka}}, \bibinfo {author} {\bibfnamefont {A.}~\bibnamefont {Keselman}},
  \bibinfo {author} {\bibfnamefont {E.}~\bibnamefont {Berg}}, \bibinfo {author}
  {\bibfnamefont {A.}~\bibnamefont {Yacoby}}, \bibinfo {author} {\bibfnamefont
  {A.}~\bibnamefont {Stern}}, \ and\ \bibinfo {author} {\bibfnamefont {B.~I.}\
  \bibnamefont {Halperin}},\ }\href {\doibase 10.1103/PhysRevX.7.021032}
  {\bibfield  {journal} {\bibinfo  {journal} {Phys. Rev. X}\ }\textbf {\bibinfo
  {volume} {7}},\ \bibinfo {pages} {021032} (\bibinfo {year}
  {2017})}\BibitemShut {NoStop}%
\bibitem [{\citenamefont {Hell}\ \emph
  {et~al.}(2017{\natexlab{b}})\citenamefont {Hell}, \citenamefont {Flensberg},\
  and\ \citenamefont {Leijnse}}]{Hell2017Coupling}%
  \BibitemOpen
  \bibfield  {author} {\bibinfo {author} {\bibfnamefont {M.}~\bibnamefont
  {Hell}}, \bibinfo {author} {\bibfnamefont {K.}~\bibnamefont {Flensberg}}, \
  and\ \bibinfo {author} {\bibfnamefont {M.}~\bibnamefont {Leijnse}},\ }\href
  {\doibase 10.1103/PhysRevB.96.035444} {\bibfield  {journal} {\bibinfo
  {journal} {Phys. Rev. B}\ }\textbf {\bibinfo {volume} {96}},\ \bibinfo
  {pages} {035444} (\bibinfo {year} {2017}{\natexlab{b}})}\BibitemShut
  {NoStop}%
\bibitem [{\citenamefont {Hart}\ \emph {et~al.}(2017)\citenamefont {Hart},
  \citenamefont {Ren}, \citenamefont {Kosowsky}, \citenamefont {Ben-Shach},
  \citenamefont {Leubner}, \citenamefont {Br{\"u}ne}, \citenamefont {Buhmann},
  \citenamefont {Molenkamp}, \citenamefont {Halperin},\ and\ \citenamefont
  {Yacoby}}]{Hart2017controlled}%
  \BibitemOpen
  \bibfield  {author} {\bibinfo {author} {\bibfnamefont {S.}~\bibnamefont
  {Hart}}, \bibinfo {author} {\bibfnamefont {H.}~\bibnamefont {Ren}}, \bibinfo
  {author} {\bibfnamefont {M.}~\bibnamefont {Kosowsky}}, \bibinfo {author}
  {\bibfnamefont {G.}~\bibnamefont {Ben-Shach}}, \bibinfo {author}
  {\bibfnamefont {P.}~\bibnamefont {Leubner}}, \bibinfo {author} {\bibfnamefont
  {C.}~\bibnamefont {Br{\"u}ne}}, \bibinfo {author} {\bibfnamefont
  {H.}~\bibnamefont {Buhmann}}, \bibinfo {author} {\bibfnamefont {L.~W.}\
  \bibnamefont {Molenkamp}}, \bibinfo {author} {\bibfnamefont {B.~I.}\
  \bibnamefont {Halperin}}, \ and\ \bibinfo {author} {\bibfnamefont
  {A.}~\bibnamefont {Yacoby}},\ }\href
  {https://www.nature.com/articles/nphys3877} {\bibfield  {journal} {\bibinfo
  {journal} {Nat. Phys.}\ }\textbf {\bibinfo {volume} {13}},\ \bibinfo {pages}
  {87} (\bibinfo {year} {2017})}\BibitemShut {NoStop}%
\bibitem [{\citenamefont {Brouwer}\ \emph
  {et~al.}(2011{\natexlab{b}})\citenamefont {Brouwer}, \citenamefont
  {Duckheim}, \citenamefont {Romito},\ and\ \citenamefont {von
  Oppen}}]{Brouwer2011topological}%
  \BibitemOpen
  \bibfield  {author} {\bibinfo {author} {\bibfnamefont {P.~W.}\ \bibnamefont
  {Brouwer}}, \bibinfo {author} {\bibfnamefont {M.}~\bibnamefont {Duckheim}},
  \bibinfo {author} {\bibfnamefont {A.}~\bibnamefont {Romito}}, \ and\ \bibinfo
  {author} {\bibfnamefont {F.}~\bibnamefont {von Oppen}},\ }\href {\doibase
  10.1103/PhysRevB.84.144526} {\bibfield  {journal} {\bibinfo  {journal} {Phys.
  Rev. B}\ }\textbf {\bibinfo {volume} {84}},\ \bibinfo {pages} {144526}
  (\bibinfo {year} {2011}{\natexlab{b}})}\BibitemShut {NoStop}%
\bibitem [{\citenamefont {Ren}\ \emph {et~al.}()\citenamefont {Ren},
  \citenamefont {Pientka}, \citenamefont {Hart}, \citenamefont {Pierce},
  \citenamefont {Kosowsky}, \citenamefont {Lunczer}, \citenamefont {Schlereth},
  \citenamefont {Scharf}, \citenamefont {Hankiewicz}, \citenamefont {Molenkamp}
  \emph {et~al.}}]{Ren2018topological}%
  \BibitemOpen
  \bibfield  {author} {\bibinfo {author} {\bibfnamefont {H.}~\bibnamefont
  {Ren}}, \bibinfo {author} {\bibfnamefont {F.}~\bibnamefont {Pientka}},
  \bibinfo {author} {\bibfnamefont {S.}~\bibnamefont {Hart}}, \bibinfo {author}
  {\bibfnamefont {A.}~\bibnamefont {Pierce}}, \bibinfo {author} {\bibfnamefont
  {M.}~\bibnamefont {Kosowsky}}, \bibinfo {author} {\bibfnamefont
  {L.}~\bibnamefont {Lunczer}}, \bibinfo {author} {\bibfnamefont
  {R.}~\bibnamefont {Schlereth}}, \bibinfo {author} {\bibfnamefont
  {B.}~\bibnamefont {Scharf}}, \bibinfo {author} {\bibfnamefont {E.~M.}\
  \bibnamefont {Hankiewicz}}, \bibinfo {author} {\bibfnamefont {L.~W.}\
  \bibnamefont {Molenkamp}},  \emph {et~al.},\ }\href
  {https://arxiv.org/abs/1809.03076} {\bibinfo  {journal} {arXiv:1809.03076}\
  }\BibitemShut {NoStop}%
\bibitem [{\citenamefont {Fornieri}\ \emph {et~al.}(2018)\citenamefont
  {Fornieri}, \citenamefont {Whiticar}, \citenamefont {Setiawan}, \citenamefont
  {Mar{\'\i}n}, \citenamefont {Drachmann}, \citenamefont {Keselman},
  \citenamefont {Gronin}, \citenamefont {Thomas}, \citenamefont {Wang},
  \citenamefont {Kallaher} \emph {et~al.}}]{Fornieri2018evidence}%
  \BibitemOpen
\bibfield  {journal} {  }\bibfield  {author} {\bibinfo {author} {\bibfnamefont
  {A.}~\bibnamefont {Fornieri}}, \bibinfo {author} {\bibfnamefont {A.~M.}\
  \bibnamefont {Whiticar}}, \bibinfo {author} {\bibfnamefont {F.}~\bibnamefont
  {Setiawan}}, \bibinfo {author} {\bibfnamefont {E.~P.}\ \bibnamefont
  {Mar{\'\i}n}}, \bibinfo {author} {\bibfnamefont {A.~C.}\ \bibnamefont
  {Drachmann}}, \bibinfo {author} {\bibfnamefont {A.}~\bibnamefont {Keselman}},
  \bibinfo {author} {\bibfnamefont {S.}~\bibnamefont {Gronin}}, \bibinfo
  {author} {\bibfnamefont {C.}~\bibnamefont {Thomas}}, \bibinfo {author}
  {\bibfnamefont {T.}~\bibnamefont {Wang}}, \bibinfo {author} {\bibfnamefont
  {R.}~\bibnamefont {Kallaher}},  \emph {et~al.},\ }\href
  {https://arxiv.org/abs/1809.03037} {\bibfield  {journal} {\bibinfo  {journal}
  {arXiv:1809.03037}\ } (\bibinfo {year} {2018})}\BibitemShut {NoStop}%
\bibitem [{\citenamefont {Lee}\ and\ \citenamefont
  {Fisher}(1981)}]{Lee1981Anderson}%
  \BibitemOpen
  \bibfield  {author} {\bibinfo {author} {\bibfnamefont {P.~A.}\ \bibnamefont
  {Lee}}\ and\ \bibinfo {author} {\bibfnamefont {D.~S.}\ \bibnamefont
  {Fisher}},\ }\href {\doibase 10.1103/PhysRevLett.47.882} {\bibfield
  {journal} {\bibinfo  {journal} {Phys. Rev. Lett.}\ }\textbf {\bibinfo
  {volume} {47}},\ \bibinfo {pages} {882} (\bibinfo {year} {1981})}\BibitemShut
  {NoStop}%
\bibitem [{SM()}]{SM}%
  \BibitemOpen
  \href@noop {} {}\bibinfo {note} {See Supplemental Material for details on:
  (i) numerical simulations, (ii) analysis of the low-energy model, and (iii)
  calculation of the localization length in the case of a single-channel
  $p$-wave and $s$-wave SCs, which include
  Refs.~\cite{Fisher1981relation,Iida1990statistical,Potter2011Majorana,Bardeen1961tunneling,Haim2016interaction,Halperin1967properties,Dorokhov1982transmission,Mello1988macroscopic}.}\BibitemShut
  {Stop}%
\bibitem [{\citenamefont {Akhmerov}\ \emph {et~al.}(2011)\citenamefont
  {Akhmerov}, \citenamefont {Dahlhaus}, \citenamefont {Hassler}, \citenamefont
  {Wimmer},\ and\ \citenamefont {Beenakker}}]{Akhmerov2011quantized}%
  \BibitemOpen
  \bibfield  {author} {\bibinfo {author} {\bibfnamefont {A.~R.}\ \bibnamefont
  {Akhmerov}}, \bibinfo {author} {\bibfnamefont {J.~P.}\ \bibnamefont
  {Dahlhaus}}, \bibinfo {author} {\bibfnamefont {F.}~\bibnamefont {Hassler}},
  \bibinfo {author} {\bibfnamefont {M.}~\bibnamefont {Wimmer}}, \ and\ \bibinfo
  {author} {\bibfnamefont {C.~W.~J.}\ \bibnamefont {Beenakker}},\ }\href
  {\doibase 10.1103/PhysRevLett.106.057001} {\bibfield  {journal} {\bibinfo
  {journal} {Phys. Rev. Lett.}\ }\textbf {\bibinfo {volume} {106}},\ \bibinfo
  {pages} {057001} (\bibinfo {year} {2011})}\BibitemShut {NoStop}%
\bibitem [{\citenamefont {Fulga}\ \emph {et~al.}(2011)\citenamefont {Fulga},
  \citenamefont {Hassler}, \citenamefont {Akhmerov},\ and\ \citenamefont
  {Beenakker}}]{Fulga2011scattering}%
  \BibitemOpen
  \bibfield  {author} {\bibinfo {author} {\bibfnamefont {I.~C.}\ \bibnamefont
  {Fulga}}, \bibinfo {author} {\bibfnamefont {F.}~\bibnamefont {Hassler}},
  \bibinfo {author} {\bibfnamefont {A.~R.}\ \bibnamefont {Akhmerov}}, \ and\
  \bibinfo {author} {\bibfnamefont {C.~W.~J.}\ \bibnamefont {Beenakker}},\
  }\href {\doibase 10.1103/PhysRevB.83.155429} {\bibfield  {journal} {\bibinfo
  {journal} {Phys. Rev. B}\ }\textbf {\bibinfo {volume} {83}},\ \bibinfo
  {pages} {155429} (\bibinfo {year} {2011})}\BibitemShut {NoStop}%
\bibitem [{\citenamefont {Beenakker}(1997)}]{Beenakker1997Random}%
  \BibitemOpen
  \bibfield  {author} {\bibinfo {author} {\bibfnamefont {C.~W.~J.}\
  \bibnamefont {Beenakker}},\ }\href {\doibase 10.1103/RevModPhys.69.731}
  {\bibfield  {journal} {\bibinfo  {journal} {Rev. Mod. Phys.}\ }\textbf
  {\bibinfo {volume} {69}},\ \bibinfo {pages} {731} (\bibinfo {year}
  {1997})}\BibitemShut {NoStop}%
\bibitem [{\citenamefont {Evers}\ and\ \citenamefont
  {Mirlin}(2008)}]{Evers2008Anderson}%
  \BibitemOpen
  \bibfield  {author} {\bibinfo {author} {\bibfnamefont {F.}~\bibnamefont
  {Evers}}\ and\ \bibinfo {author} {\bibfnamefont {A.~D.}\ \bibnamefont
  {Mirlin}},\ }\href {\doibase 10.1103/RevModPhys.80.1355} {\bibfield
  {journal} {\bibinfo  {journal} {Rev. Mod. Phys.}\ }\textbf {\bibinfo {volume}
  {80}},\ \bibinfo {pages} {1355} (\bibinfo {year} {2008})}\BibitemShut
  {NoStop}%
\bibitem [{Note2()}]{Note2}%
  \BibitemOpen
  \bibinfo {note} {We limit $U_{m}(\protect \boldsymbol {r})$ to the junction
  since we are not interested here in its effect on the SCs.}\BibitemShut
  {Stop}%
\bibitem [{\citenamefont {Fisher}\ and\ \citenamefont
  {Lee}(1981)}]{Fisher1981relation}%
  \BibitemOpen
  \bibfield  {author} {\bibinfo {author} {\bibfnamefont {D.~S.}\ \bibnamefont
  {Fisher}}\ and\ \bibinfo {author} {\bibfnamefont {P.~A.}\ \bibnamefont
  {Lee}},\ }\href {\doibase 10.1103/PhysRevB.23.6851} {\bibfield  {journal}
  {\bibinfo  {journal} {Phys. Rev. B}\ }\textbf {\bibinfo {volume} {23}},\
  \bibinfo {pages} {6851} (\bibinfo {year} {1981})}\BibitemShut {NoStop}%
\bibitem [{\citenamefont {Iida}\ \emph {et~al.}(1990)\citenamefont {Iida},
  \citenamefont {Weidenm\"uller},\ and\ \citenamefont
  {Zuk}}]{Iida1990statistical}%
  \BibitemOpen
  \bibfield  {author} {\bibinfo {author} {\bibfnamefont {S.}~\bibnamefont
  {Iida}}, \bibinfo {author} {\bibfnamefont {H.~A.}\ \bibnamefont
  {Weidenm\"uller}}, \ and\ \bibinfo {author} {\bibfnamefont {J.}~\bibnamefont
  {Zuk}},\ }\href
  {http://www.sciencedirect.com/science/article/pii/000349169090275S}
  {\bibfield  {journal} {\bibinfo  {journal} {Ann. Phys.}\ }\textbf {\bibinfo
  {volume} {200}},\ \bibinfo {pages} {219 } (\bibinfo {year}
  {1990})}\BibitemShut {NoStop}%
\bibitem [{\citenamefont {Potter}\ and\ \citenamefont
  {Lee}(2011)}]{Potter2011Majorana}%
  \BibitemOpen
  \bibfield  {author} {\bibinfo {author} {\bibfnamefont {A.~C.}\ \bibnamefont
  {Potter}}\ and\ \bibinfo {author} {\bibfnamefont {P.~A.}\ \bibnamefont
  {Lee}},\ }\href {\doibase 10.1103/PhysRevB.83.094525} {\bibfield  {journal}
  {\bibinfo  {journal} {Phys. Rev. B}\ }\textbf {\bibinfo {volume} {83}},\
  \bibinfo {pages} {094525} (\bibinfo {year} {2011})}\BibitemShut {NoStop}%
\bibitem [{\citenamefont {Bardeen}(1961)}]{Bardeen1961tunneling}%
  \BibitemOpen
  \bibfield  {author} {\bibinfo {author} {\bibfnamefont {J.}~\bibnamefont
  {Bardeen}},\ }\href {\doibase 10.1103/PhysRevLett.6.57} {\bibfield  {journal}
  {\bibinfo  {journal} {Phys. Rev. Lett.}\ }\textbf {\bibinfo {volume} {6}},\
  \bibinfo {pages} {57} (\bibinfo {year} {1961})}\BibitemShut {NoStop}%
\bibitem [{\citenamefont {Haim}\ \emph {et~al.}(2016)\citenamefont {Haim},
  \citenamefont {W\"olms}, \citenamefont {Berg}, \citenamefont {Oreg},\ and\
  \citenamefont {Flensberg}}]{Haim2016interaction}%
  \BibitemOpen
  \bibfield  {author} {\bibinfo {author} {\bibfnamefont {A.}~\bibnamefont
  {Haim}}, \bibinfo {author} {\bibfnamefont {K.}~\bibnamefont {W\"olms}},
  \bibinfo {author} {\bibfnamefont {E.}~\bibnamefont {Berg}}, \bibinfo {author}
  {\bibfnamefont {Y.}~\bibnamefont {Oreg}}, \ and\ \bibinfo {author}
  {\bibfnamefont {K.}~\bibnamefont {Flensberg}},\ }\href {\doibase
  10.1103/PhysRevB.94.115124} {\bibfield  {journal} {\bibinfo  {journal} {Phys.
  Rev. B}\ }\textbf {\bibinfo {volume} {94}},\ \bibinfo {pages} {115124}
  (\bibinfo {year} {2016})}\BibitemShut {NoStop}%
\bibitem [{\citenamefont {Halperin}(1967)}]{Halperin1967properties}%
  \BibitemOpen
  \bibfield  {author} {\bibinfo {author} {\bibfnamefont {B.~I.}\ \bibnamefont
  {Halperin}},\ }\href
  {https://onlinelibrary.wiley.com/doi/abs/10.1002/9780470140154.ch6} {\emph
  {\bibinfo {title} {Properties of a Particle in a One-Dimensional Random
  Potential}}},\ edited by\ \bibinfo {editor} {\bibfnamefont {I.}~\bibnamefont
  {Prigogine}},\ Vol.~\bibinfo {volume} {13}\ (\bibinfo  {publisher} {Wiley
  Online Library},\ \bibinfo {year} {1967})\ pp.\ \bibinfo {pages}
  {123--177}\BibitemShut {NoStop}%
\bibitem [{\citenamefont {Dorokhov}(1982)}]{Dorokhov1982transmission}%
  \BibitemOpen
  \bibfield  {author} {\bibinfo {author} {\bibfnamefont {O.}~\bibnamefont
  {Dorokhov}},\ }\href {http://www.jetpletters.ac.ru/ps/1335/article_27160.pdf}
  {\bibfield  {journal} {\bibinfo  {journal} {JETP Lett}\ }\textbf {\bibinfo
  {volume} {36}},\ \bibinfo {pages} {318} (\bibinfo {year} {1982})}\BibitemShut
  {NoStop}%
\bibitem [{\citenamefont {Mello}\ \emph {et~al.}(1988)\citenamefont {Mello},
  \citenamefont {Pereyra},\ and\ \citenamefont {Kumar}}]{Mello1988macroscopic}%
  \BibitemOpen
  \bibfield  {author} {\bibinfo {author} {\bibfnamefont {P.}~\bibnamefont
  {Mello}}, \bibinfo {author} {\bibfnamefont {P.}~\bibnamefont {Pereyra}}, \
  and\ \bibinfo {author} {\bibfnamefont {N.}~\bibnamefont {Kumar}},\ }\href
  {https://www.sciencedirect.com/science/article/pii/0003491688901698}
  {\bibfield  {journal} {\bibinfo  {journal} {Ann. Phys.}\ }\textbf {\bibinfo
  {volume} {181}},\ \bibinfo {pages} {290} (\bibinfo {year}
  {1988})}\BibitemShut {NoStop}%
\end{thebibliography}%

\newpage
\begin{widetext}
	
\section*{Supplemental Material}

\section{Details of numerical simulations}

In this section we present details of the numerical simulations whose
results are summarized in Fig. 1 of the main text. We begin by presenting
the lattice model used for simulating the system. We then explain
the procedure for obtaining the reflection matrix and extracting the Majorana localization length.

\subsection{The lattice model}

For the purpose of numerically simulating the planar Josephson junction
[Eq. (1) of the main text], we replace it with a
model of a $N_{x}\times N_{y}$ square lattice of lattice constant
$a$, whose Hamiltonian is given by
\begin{equation}
\begin{split}H_{{\rm PJJ}}=\sum_{n_{x}=1}^{N_{x}}\sum_{n_{y}=1}^{N_{y}}\sum_{s,s'\in\{\uparrow,\downarrow\}} & \Big\{[(U_{n_{x},n_{y}}-\mu_{n_{y}})\sigma_{ss'}^{0}-E_{n_{y}}^{{\rm Z}}\sigma_{ss'}^{x}]c_{\boldsymbol{n},s}^{\dagger}c_{\boldsymbol{n},s'}-\sum_{\boldsymbol{d}\in\{\pm\hat{x},\pm\hat{y\}}}[t_{0}\sigma_{ss'}^{0}+iu(\boldsymbol{\sigma}_{ss'}\times\boldsymbol{d})\cdot\hat{z}]c_{\boldsymbol{n},s}^{\dagger}c_{\boldsymbol{n}+\boldsymbol{d},s'}\\
& +\frac{1}{2}[\Delta_{n_{y}}i\sigma_{ss'}^{y}c_{\boldsymbol{n},s}^{\dagger}c_{\boldsymbol{n},s'}^{\dagger}+{\rm h.c.}]\Big\}
\end{split}
\end{equation}
where $c_{\boldsymbol{n},s}^{\dagger}$ creates an electron on site
$\boldsymbol{n}=(n_{x},n_{y})$, $U_{n_{x},n_{y}}=U(n_{x}a,n_{y}a)$,
$\mu_{n_{y}}=\mu(n_{y}a)-4t_0$, $E_{n_{y}}^{{\rm Z}}=E_{{\rm Z}}(n_{y}a)$,
$\Delta_{n_{y}}=\Delta(n_{y}a)$, $t_{0}=1/2m_{{\rm e}}a^{2}$, $u=\alpha/2a$,
$N_{x}=L_{x}/a$, and $N_{y}=(2W_{{\rm sc}}+W)/a$. In the present
work, we use $t_{0}=2.5$.

\subsection{The reflection matrix}\label{sec:ref_mat}

We begin by rewriting the Hamiltonian in the following form
\begin{equation}
\begin{split} &
H_{{\rm PJJ}}=\sum_{n_{x}=1}^{N_x}\vec{\psi}_{n_{x}}^{\dagger}h_{n_{x}}\vec{\psi}_{n_{x}}+\left[\vec{\psi}_{n_{x}}^{\dagger}V\vec{\psi}_{n_{x}+1}+{\rm h.c.}\right],
\end{split}
\end{equation}
where $\vec{\psi}_{n_{x}}^{\dagger}=(c_{n_{x},1,\uparrow}^{\dagger},c_{n_{x},1,\uparrow},c_{n_{x},1,\downarrow}^{\dagger},c_{n_{x},1,\downarrow},\dots,c_{n_{x},N_{y},\uparrow}^{\dagger},c_{n_{x},N_{y},\uparrow},c_{n_{x},N_{y},\downarrow}^{\dagger},c_{n_{x},N_{y},\downarrow})$
is a $1\times4N_{y}$ vector of creation and annihilation operators, and where $\{h_{n_{x}}\}_{n_{x}=1}^{N_{x}}$
and $V$ are $4N_{y}\times4N_{y}$ matrices.

We place two normal-metal leads, at $x=0$ and $x=L_x$. The reflection matrix for electrons and holes incident from the right is given by~\cite{Fisher1981relation,Iida1990statistical}
\begin{eqnarray}
r(\omega) = \mathbbm{1}-2\pi iW_{{\rm R}}^{\dagger}[G_{N_{x}}^{-1}(\omega)+i\pi W_{{\rm R}}W_{{\rm R}}^{\dagger}]W_{{\rm R}},
\end{eqnarray}
where $W_{\ensuremath{{\rm R}}}\equiv \sqrt{\rho_{\rm R}}V$, with $\rho_{\rm R}$ being the density of states in the right lead, and $G_{N_x}$ is the Green function matrix at the right-most sites of the system, obtained through the recursive relation~\cite{Lee1981Anderson}
\begin{equation}\label{eq:Local_Green}
G_{n_{x}}(\omega)=[\omega-h_{n_{x}}-V^{\dagger}G_{n_{x}-1}V]^{-1}.
\end{equation}
Here, $G_{n_x}(\omega)$ is a $4N_y\times4N_y$ matrix for every $n_x$ (indices running over spin, particle-hole and $n_y$), and $G_0=-i\pi\rho_{\rm L}$, with $\rho_{\rm L}$ being the density of states in the left lead.

\subsection{Topological invariant and localization length}

Given the reflection matrix, the topological invariant is given by~\cite{Akhmerov2011quantized,Fulga2011scattering}
$\mathcal{Q}=\lim_{N_{x}\to\infty}\det[r(0)]$, which takes the value
$+1$ ($-1$) in the trivial (topological) phase. As an example, in Fig.~\hyperref[fig:Q_n_and_T_n]{\ref{fig:Q_n_and_T_n}(a)} we present $\det[r(0)]$ as a function of system's length, $N_{x}$, for four different disorder realizations, with increasing value of disorder
strength. The rest of the system parameters are as in Fig. 1 of the
main text, with $E_{{\rm Z,J}}=1$ and $\phi=\pi$. When calculating the topological invariant for a clean system [Fig.~1(a) of the main text] we have instead used the Pfaffian invariant introduced in Ref.~\cite{Kitaev2001unpaired}. 

To obtain the localization length, the transmission probability matrix
is obtained through $T(\omega)\equiv t^{\dagger}(\omega)t(\omega)=1-r^{\dagger}(\omega)r(\omega)$,
where $t(\omega)$ is the transmission matrix, and we used the fact that the scattering matrix is unitary. The Majorana localization length is determined by the decay
of the largest eigenvalue of $T(0).$ This eigenvalue is shown in
Fig.~\hyperref[fig:Q_n_and_T_n]{\ref{fig:Q_n_and_T_n}(b)} as a function of $N_{x}$ for four
different value of disorder strength. We then extract the localization
length by computing
\begin{equation}
\xi=a\sum_{N_{x}=1}^{N_x^{\rm max}\to\infty}T_{{\rm max}}(\omega=0,N_{x}a),
\end{equation}
where by $T_{{\rm max}}(\omega,L_{x})$ we denote the largest eigenvalue
of the transmission probability, for a system of length $L_{x}.$
Notice that for an exponentially decaying transmission, $T_{{\rm max}}(0,L_{x})=\exp(-\lambda L_{x})$,
this indeed yields the decay length, $\xi=1/\lambda,$ assuming the
lattice spacing is taking to be small enough $(a\ll\xi)$. In the
simulations presented in this work we averaged $\xi$
over a 100 realizations for every data point, and the maximal system's length was $N^{\rm max}_x=10^4$.

Finally, in Fig.~\hyperref[fig:Q_n_and_T_n]{\ref{fig:Q_n_and_T_n}(c)} we present the $x$ profile of the zero-energy local density of states,  $\rho(x)\equiv\int{\rm d}y\mathcal{N}(\omega=0,x,y)$, for the four disorder realizations corresponding to Figs.~\hyperref[fig:Q_n_and_T_n]{\ref{fig:Q_n_and_T_n}(a)} and ~\hyperref[fig:Q_n_and_T_n]{\ref{fig:Q_n_and_T_n}(b)}. The local density of states, $\mathcal{N}(\omega,x,y)$, was calculated according to the method described in Ref.~\cite{Potter2011Majorana}.

\begin{figure}
	\begin{centering}
		\begin{tabular}{ccc}
			\includegraphics[clip=true,trim =0mm 0mm 0mm 0mm,height=4.5cm]{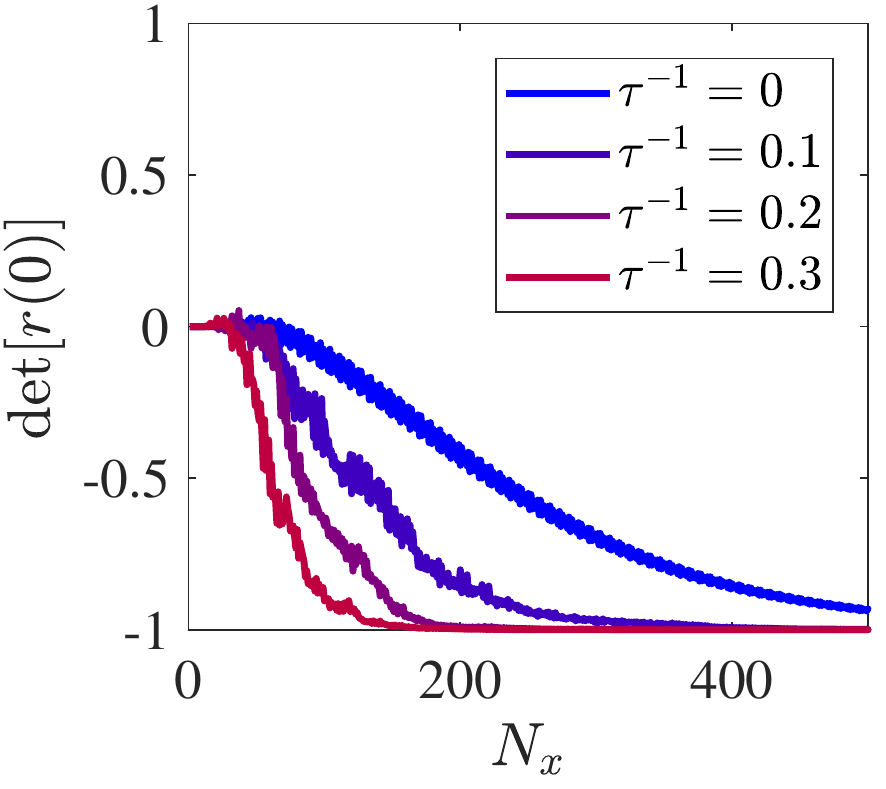}
			\llap{\hskip 0mm \parbox[c]{8.5cm}{\vspace{0cm}(a)}}&
			\includegraphics[clip=true,trim =0mm 3mm 0mm 0mm,height=4.4cm]{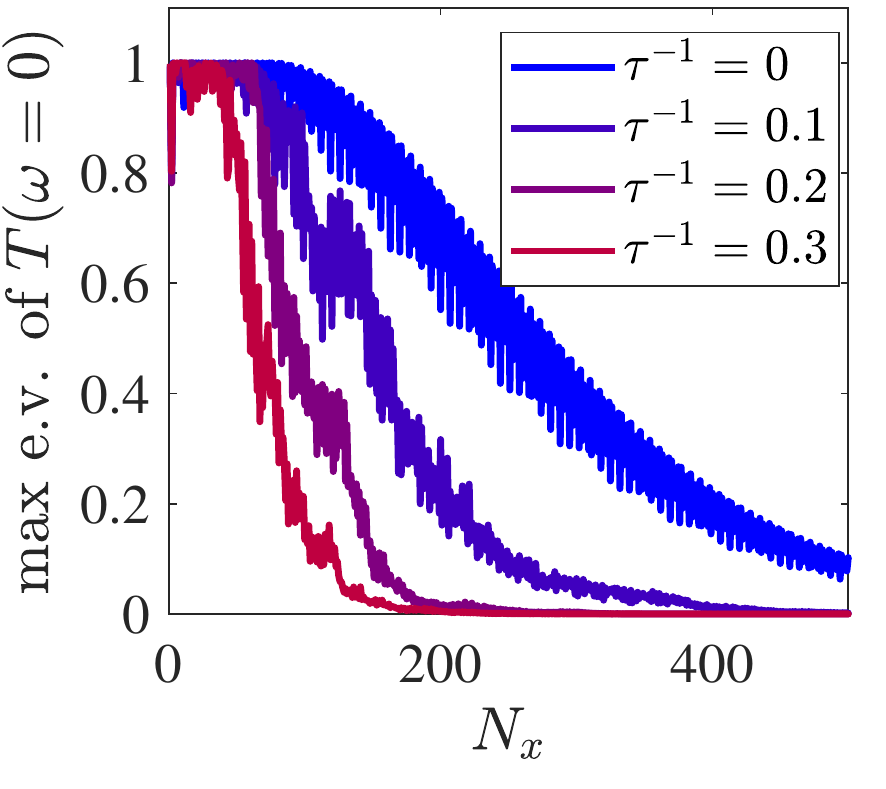}
			\llap{\hskip 0mm \parbox[c]{8.8cm}{\vspace{0cm}(b)}}&
			\includegraphics[clip=true,trim =0mm 0mm 0mm 0mm,height=4.5cm]{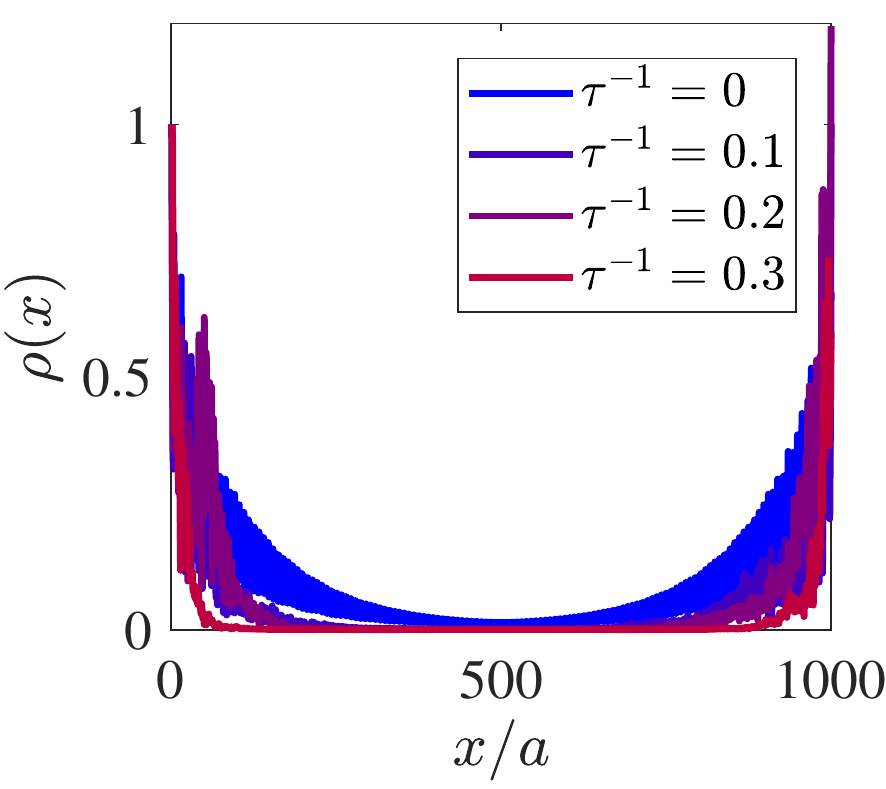}
			\llap{\hskip 0mm \parbox[c]{8.6cm}{\vspace{0cm}(c)}}
			\tabularnewline
		\end{tabular}
		\par\end{centering}
	\caption{(a) The topological invariant, $\mathcal{Q}=\det[r(\omega=0)]$ as
		a function of the system's length, $N_{x}=L_{x}/a$, for different
		disorder strength, characterized by the inverse mean free time of
		the bare 2DEG, $\tau^{-1}$. For all the disorder strength presented, $\tau^{-1}=0,0.1,0.2,0.3$
		the system is in the topological phase. Each plot here corresponds to a single disorder realization. (b) The maximal eigenvalue of
		the zero-frequency transmission probability matrix, $T(\omega=0)$,
		for the three disorder strength values in the topological phase. (c) The profile of the local density of states, integrated along the $y$ direction, $\rho(x)\int{\rm d}y\mathcal{N}(\omega=0,x,y)$, for the different disorder strengths. The
		system's parameter are the same as in Fig. (1) of the main text,
		$E_{{\rm Z,J}}=1$ and $\phi=\pi$. \label{fig:Q_n_and_T_n}}
\end{figure}

\subsection{results for different parameters}

In Fig.~\ref{fig:PJJ_SM}, we present results for a junction with parameters different from those shown in Fig.~1 of the main text. Figure~\hyperref[fig:PJJ_SM]{\ref{fig:PJJ_SM}(a)} presents the phase diagram in the clean limit, and Fig.~\hyperref[fig:PJJ_SM]{\ref{fig:PJJ_SM}(b)} presents the Majorana localization  length as a function of disorder strength, for chemical potential $\mu_{\rm J}=\mu_{\rm sc}=0.5$. The rest of the parameters are the same as in Fig.~1 of the main text. The same qualitative behavior is observed as in Fig.~1 of the main text.

In Fig.~\hyperref[fig:PJJ_SM]{\ref{fig:PJJ_SM}(c)} we examine the effect of disorder on the system's phase diagram, for $\mu_{\rm J}=\mu_{\rm sc}=1$, and for a narrower junction, $W=2.5l_{\rm so}$ (compared with $W=5l_{\rm so}$ in the main text). The rest of the system parameters are the same as in Fig.~1 of the main text. The topological invariant, $\mathcal{Q}=\det[r(0)]$, is shown for a disorder strength of $\tau^{-1}=0.5$. The black dashed line represents the phase boundaries in the case of the clean system with the same parameters. Interestingly, for some magnetic fields, $E_{\rm Z,J}$, and phase biases, $\phi$, disorder drives the system from the trivial phase to the topological phase. A similar effect was previously observed in Refs.~\cite{Adagideli2014effects,Pientka2013signatures}.

\begin{figure}
	\begin{tabular}{lcr}
		\hskip -0mm
		\includegraphics[clip,trim= 0 0 0 2mm,height=4.5cm]{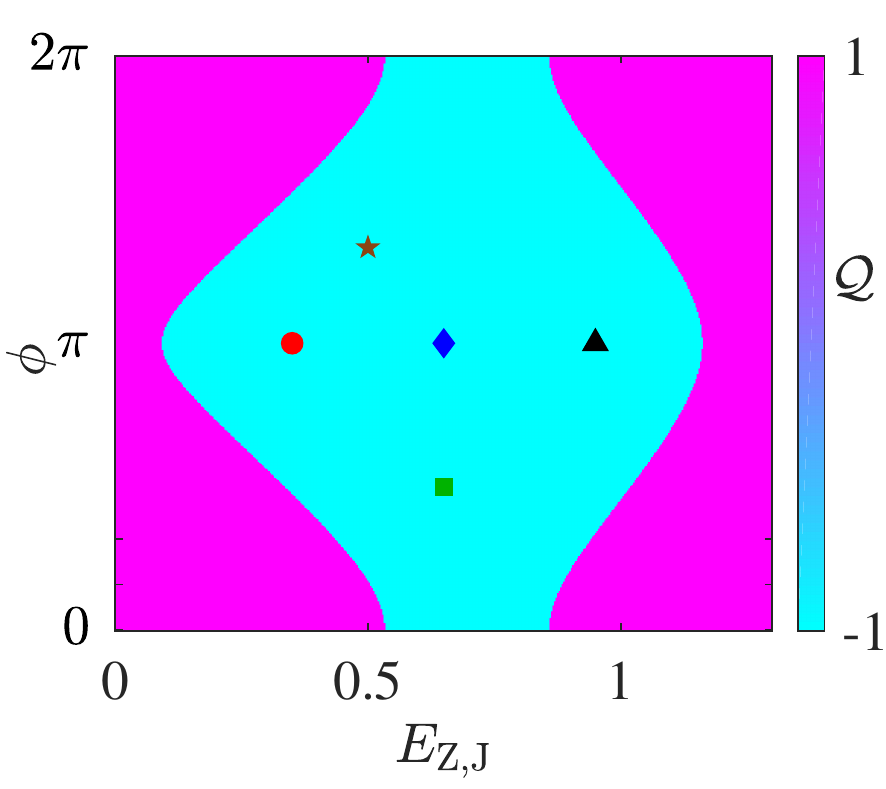}
		\llap{\hskip 2mm \parbox[c]{10cm}{\vspace{-3mm}(a)}}
		\hskip 0mm
		&
		\includegraphics[clip,trim = 0mm 0mm 0mm 0mm,height=4.4cm]{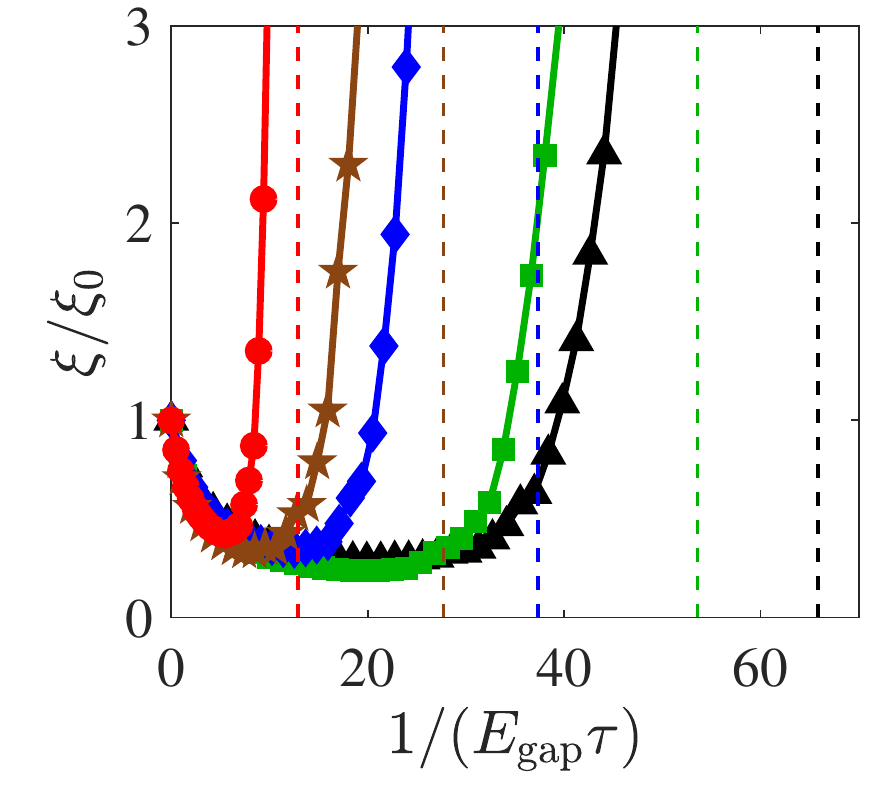}
		\llap{\hskip 2mm \parbox[c]{9cm}{\vspace{-3mm}(b)}}
		&
		\hskip 2mm
		\includegraphics[clip,trim= 0 0 0 2mm,height=4.5cm]{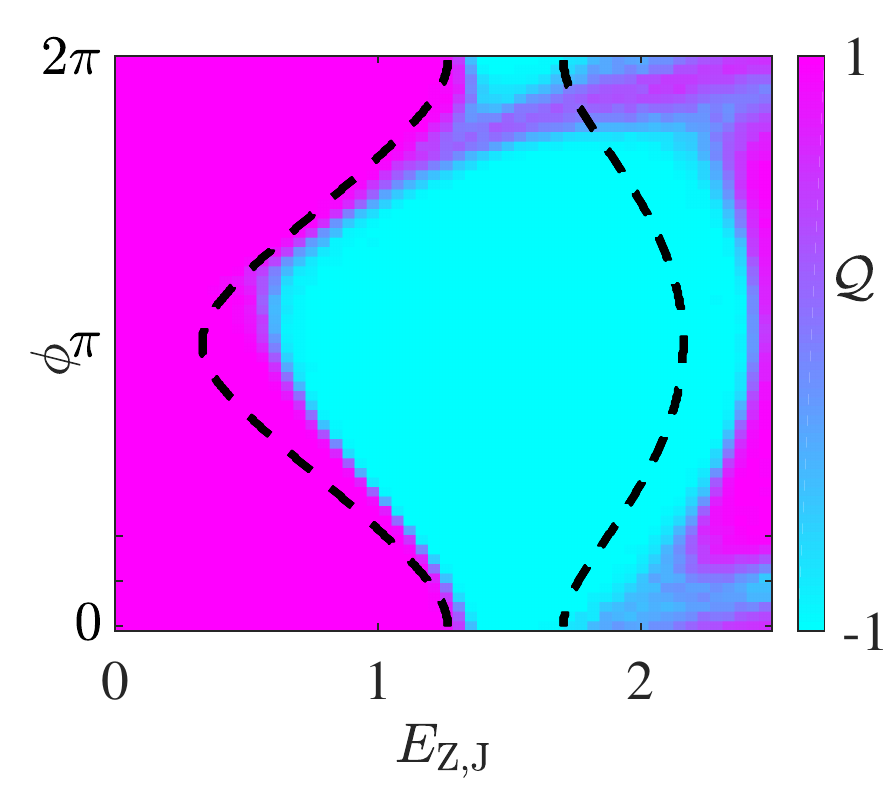}
		\llap{\hskip 2mm \parbox[c]{10cm}{\vspace{-3mm}(c)}}
	\end{tabular}
	\vskip -3mm
	\caption{Results for different parameters. (a) The topological (blue) and trivial (pink) regions in the $\phi$--$B$ plane for a clean system, for $E_{{\rm so}}=m_{\rm e}\alpha^2/2=1$, $\Delta_0=1$, $\mu_{{\rm J}}=\mu_{{\rm SC}}=0.5$, $E_{\rm Z,SC}=0$, $l_{{\rm so}}=1/m_{\rm e}\alpha=0.2W$, $W_{\rm sc}=W$ (b) The Majorana localization length, $\xi$, as a function
		of the disorder-induced inverse mean free time, $\tau^{-1}$, for different points
		inside the topological phase [see markers in (a)]. In (b), $\tau$ is normalized by the overall gap in the clean system, $E_{\rm gap}$, which is $0.035$, $0.033$, $0.039$, $0.022$, and $0.029$ for the red, brown, blue, green and black plots, respectively. (c) Phase diagram for chemical potential $\mu_{\rm J}=\mu_{\rm SC}=1$, and junction width, $W=2.5l_{\rm so}$, for disorder strength, $\tau^{-1}=0.5$. The reset of the system parameters are unchanged. The phase boundaries in the case of the corresponding clean system are shown in a black dashed line.
		\label{fig:PJJ_SM}}
\end{figure}

\section{The analysis of the linearized multi-channel model}

In the main text we have studied a linearized low-energy model describing
a disordered multi-channel superconductor, Eq.~(2), and
performed a perturbative analysis of the disorder, which resulted
in new effective pairing potentials, Eq.~(5).
In this section we explain how this model can arise from a microscopic
model, such as the planar JJ, Eq.~(1) of the main text, and provide details
regarding the calculation of the self energy which yielded the expression
for the effective pairing potentials.

\subsection{Origin of the model}

We start from the 2d model of Eq.~(1) of the main text,
and separate the system to two parts: the normal part which is the
strip defined by $|y|<W/2$, and the superconducting part, $|y|>W/2$.
Following the Bardeen tunneling-Hamiltonian approach~\cite{Bardeen1961tunneling}, we
then write the overall Hamiltonian as a combination of the three terms,
describing the normal part, the SC part, and the coupling between
them,
\begin{equation}
H=H_{{\rm N}}+H_{{\rm SC}}+H_{{\rm N-SC}},
\end{equation}
where $H_{{\rm N}}$ ($H_{{\rm SC}}$) is the Hamiltonian obtained
by imposing hard-wall boundary conditions for $|y|>W/2$ ($|y|<W/2$).
This treatment is valid when the normal reflection at the N-S interfaces
($y=\pm W/2$) is strong, such that the normal part as weakly coupled
to the SC. This is the case, in particular, for the high-momentum
modes as they impinge upon the N-S interface at large angles. Regardless
of the above considerations, our numerical analysis shows that the
qualitative conclusions drawn from the low-energy model of Eq.~(2) of the main text
hold much more generally.

We write the normal part, $H_{{\rm N}}$, as a combination as two
terms,
\begin{equation}\label{eq:_H_N}
H_{{\rm N}}=\int{\rm d}x\int_{-W/2}^{W/2}{\rm d}y\Psi^{\dagger}(x,y)\left[\mathcal{H}_{{\rm N}}^{0}(x,y)+\mathcal{H}_{{\rm N}}^{{\rm dis}}(x,y)\right]\Psi(x,y),
\end{equation}
where $\mathcal{H}_{{\rm N}}^{0}$ describes the system in the clean
limit, and $\mathcal{H}_{{\rm N}}^{{\rm dis}}=U(x,y)\tau_{z}$ is
the part coming from disorder. Our treatment of the system is composed
of two steps: (i) first we solve for $\mathcal{H}_{{\rm N}}^{0}$,
and (ii) the disorder term and the induced superconductivity are then
projected onto the basis diagonalizing $\mathcal{H}_{{\rm N}}^{0}$.

The clean part of the Hamiltonian, $\mathcal{H}_{{\rm N}}^{0}$, is
generally solved by a set of eigenstates,
\begin{equation}
\vec{\varphi}_{\nu,k_x}(x,y)=\frac{e^{ik_xx}}{\sqrt{2\pi}}\cdot\begin{pmatrix}\eta_{\nu,k_x}^{\uparrow}(y)\\
\eta_{\nu,k_x}^{\downarrow}(y)
\end{pmatrix},
\end{equation}
with corresponding eigen-energies, $E_{\nu,k_x}$, and $\nu=1,2,\dots,\infty.$
Here, $k_x$ is the momentum in the $x$ direction, while $\nu$ labels
the transverse channels.

\subsubsection{Reflection Symmetry}

The clean part of the Hamiltonian obeys the following symmetry,
\begin{equation}
\sigma_{x}\mathcal{H}_{{\rm N}}^{0}(-x,y)\sigma_{x}=\mathcal{H}_{{\rm N}}^{0}(x,y),
\end{equation}
as can be checked by setting $U(x,y)=0$ in Eq.~(1)
of the main text. The eigenstates can therefore be chosen to obey
\begin{equation}
\eta_{\nu,-k_x}^{s}(y)=\sum_{s'=\uparrow,\downarrow}\sigma_{ss'}^{x}\eta_{\nu,k_x}^{s'}(y).\label{eq:ref_sym}
\end{equation}

\subsubsection{Conducting channels}

Depending on the chemical potential, some of the bands labeled by
$\nu$ will cross zero energy, $E_{\nu,k_x}=0$, for some momentum $k_x$.
Due to the above reflection symmetry, these momenta will come in opposite-momentum
pairs (except for potentially a single Fermi point at , which can
occur when the chemical potential is at the bottom of one of the bands).
The number of bands crossing zero energy, $N$, defines the number
of conducting channels in the model. Correspondingly, we label the
Fermi momenta by $k_{{\rm F},n}$, where $n=\pm1,\dots,\pm N$, and
where $k_{{\rm F},-n}=-k_{{\rm F},n}$. Below we will be interested
only in the modes having momentum near $k_{{\rm F},n}$.

\subsubsection{Projection and Linearization}

We first project the disorder part of the Hamiltonian onto the new
basis. To this end, we first make the transformation
\begin{equation}
\hat{\psi_{s}}(x,y)=\int{\rm d}ke^{ik_xx}\sum_{\nu=1}^{\infty}\eta_{\nu,k_x}^{s}(y)\hat{a}_{\nu,k_x},
\end{equation}
where by definition, $\hat{a}_{\nu,k_x}$ creates an electron in the
state described by $\vec{\varphi}_{\nu,k_x}(x,y)$. Setting in Eq.~\eqref{eq:_H_N}, one then has
\begin{equation}
\begin{split}H_{{\rm N}}=\sum_{\nu,k_x}E_{\nu,k_x}\hat{a}_{n,k_x}^{\dagger}\hat{a}_{n,k_x}+\int{\rm d}x\sum_{k_x,k_x'}e^{i(k_x'-k_x)x}\sum_{\nu,\nu'=1}^{\infty}\sum_{s}\int_{-W/2}^{W/2}{\rm d}yU(x,y)[\eta_{\nu,k_x}^{s}(y)]^{\ast}\eta_{\nu',k_x'}^{s}(y)\hat{a}_{\nu,k_x}^{\dagger}\hat{a}_{\nu',k_x'}.
\end{split}
\label{eq:H_n_basis}
\end{equation}

Since we are concerned only with the low-energy modes, we can project
out all the bands not crossing the Fermi energy. Out of the sum over
$\nu$, this leaves us only with a sum over $n=\pm1,\dots,\pm N.$
Furthermore, we can limit the integral over $k_x$ to momenta close
to the Fermi points, $k_{{\rm F},n}$. This is done by defining the
fields living close to the Fermi momenta, $\hat{a}_{n,k_{{\rm F},n}+q}\equiv\hat{\phi}_{n,q},$
where $q\in[-\Lambda,\Lambda].$ Finally, if the bottom of all the
bands is far enough from the Fermi energy (which we shall assume to
be the case), then we can approximate the dispersions of the modes
near the Fermi points by $E_{n,k_x}\simeq\partial_{k_x}E_{n,k_x}|_{k_{{\rm F},n}}\cdot (k_x-k_{{\rm F},n})\equiv v_{n}(k_x-k_{{\rm F},n})$,
and take $\Lambda\to\infty$. Note also that due to the symmetry,
Eq. (\ref{eq:ref_sym}), one has $v_{-n}=-v_{n}.$ Applying the above
procedure to Eq. (\ref{eq:H_n_basis}), one has
\begin{equation}
\begin{split}
H_{{\rm N}}\simeq & \sum_{n=\pm1}^{\pm N}\int{\rm d}qv_{n}q\hat{\phi}_{n,q}^{\dagger}\hat{\phi}_{n,q}+\int{\rm d}x\int{\rm d}q\int{\rm d}q'e^{i(q'-q)x}\sum_{m,n=\pm1}^{\pm N}\sum_{s}\int_{-W/2}^{W/2}{\rm d}yU(x,y)[\eta_{m,k_{{\rm F},m}}^{s}(y)]^{\ast}\eta_{n,k_{{\rm F},n}}^{s}(y)\hat{\phi}_{m,q}^{\dagger}\phi_{n,q'}\\
= & \sum_{n=\pm1}^{\pm N}\int{\rm d}xv_{n}\hat{\phi}_{n}^{\dagger}(x)(-iv_{n}\partial_{x})\hat{\phi}_{n}(x)+\int{\rm d}x\sum_{m,n=\pm1}^{\pm N}V_{mn}(x)\hat{\phi}_{m}^{\dagger}(x)\hat{\phi}_{n}(x),
\end{split}
\label{eq:H_N_lin}
\end{equation}
where we have defined
\begin{equation}\label{eq:V_mn_def}
V_{mn}(x)=\sum_{s}\int_{-W/2}^{W/2}{\rm d}yU(x,y)[\eta_{m,k_{{\rm F},m}}^{s}(y)]^{\ast}\eta_{n,k_{{\rm F},n}}^{s}(y).
\end{equation}

Finally, we account for the coupling to the superconducting region.
At least in principle, one can integrate out the degrees of freedom
of the SC region~\cite{Alicea2012,Haim2016interaction}. This will result in \emph{induced} pairing
potential operating on the modes living in the normal region,
\begin{equation}
H_{{\rm N}}^{{\rm ind}}=\int{\rm d}x\sum_{m,n=\pm1}^{\pm N}\Delta_{mn}\hat{\phi}_{m}^{\dagger}(x)\hat{\phi}_{n}^{\dagger}(x).
\end{equation}
Importantly, only pairing between modes of opposite momenta will open a gap at the Fermi energy. Assuming
the Fermi momenta, $k_{{\rm F},m}$, are not degenerate (this will
generally be the case when breaking ${\rm SU}(2)$ symmetry), we can therefore omit all the pairing terms except for $\Delta_{m,-m}\equiv\Delta_{m}$. Combining
with Eq. (\ref{eq:H_N_lin}), the Hamiltonian describing the overall
system at low energies is given by
\begin{equation}
H\simeq\int{\rm d}x\left\{ \sum_{m=\pm1}^{\pm N}\left[v_{m}\hat{\phi}_{m}^{\dagger}(x)(-iv_{m}\partial_{x})\hat{\phi}_{m}(x)+\Delta_{m}\hat{\phi}_{m}^{\dagger}(x)\hat{\phi}_{-m}^{\dagger}(x)\right]+\sum_{m,n=\pm1}^{\pm N}V_{mn}(x)\hat{\phi}_{m}^{\dagger}(x)\hat{\phi}_{n}(x)\right\},
\end{equation}
which is the Hamiltonian introduced in Eq.~(2) of the main
text.

\subsubsection{Properties of the disorder term}

The new effectively-1D disorder potential, $V_{mn}(x)$, is manifestly
Hermitian, $V_{mn}^{\ast}(x)=V_{nm}(x)$. Furthermore, due to the
symmetry, Eq. (\ref{eq:ref_sym}), it obeys $V_{-m,-n}(x)=V_{mn}(x)$. From Eq.~\eqref{eq:V_mn_def}, we can obtain the correlations of $V_{mn}(x)$, which are given by
\begin{equation}
\begin{split}
\left\langle V_{mn}(x)V_{mn}(x')\right\rangle =\gamma_{mn}\delta(x-x'),
\end{split}
\label{eq:V_mnV_mn}
\end{equation}
where we defined
\begin{equation}
\gamma_{mn}\equiv\gamma\int_{-W/2}^{W/2}{\rm d}y\left[\vec{\eta}_{m,k_{{\rm F},m}}^{\dagger}(y)\vec{\eta}_{n,k_{{\rm F},n}}(y)\right]^{2},
\end{equation}
and where we have used the fact that $\langle U(x,y)U(x',y')\rangle=\gamma\delta(\boldsymbol{r}-\boldsymbol{r}')$.
Notice also that
\begin{align}
\left\langle V_{mn}(x)V_{nm}(x')\right\rangle  & =|\gamma_{mn}|\delta(x-x').\label{eq:V_mnV_nm}
\end{align}

\subsection{Derivation of the self energy}

\subsubsection{Gauging out the diagonal scattering terms}

Starting from the low-energy Hamiltonian, Eq.~(2) of the
main text, we define the fields
\begin{equation}
\tilde{\phi}_{m}(x)=\phi_{m}(x)e^{\frac{i}{v_{m}}\int_{-\infty}^{x}{\rm {\rm d}}x_{1}V_{mm}(x_{1})}.
\end{equation}
Inserting this definition into Eq.~(2) of the
main text, we arrive at an
identical Hamiltonian, except that now intra-mode scattering is absent,
\begin{equation}
\begin{split}H=\int{\rm d}x\Bigg( & \sum_{m=\pm1}^{\pm N}\left\{ v_{m}\tilde{\phi}_{m}^{\dagger}(x)(-i\partial_{x})\tilde{\phi}_{m}(x)+\frac{1}{2}\left[\Delta_{m}\tilde{\phi}_{m}^{\dagger}(x)\tilde{\phi}_{-m}^{\dagger}(x)+{\rm h.c.}\right]\right\} +\sum_{m,n=\pm1}^{\pm N}e^{i(k_{{\rm F},m}-k_{{\rm F},n})x}\tilde{V}_{mn}(x)\phi_{m}^{\dagger}(x)\phi_{n}(x)\Bigg),\end{split}
\label{eq:H_tilde_fields}
\end{equation}
where,
\begin{align}
\tilde{V}_{mn}(x)\equiv\begin{cases}
V_{mn}(x)e^{i\int_{-\infty}^{x}{\rm {\rm d}}x_{1}\left[\frac{1}{v_{m}}V_{mm}(x_{1})-\frac{1}{v_{n}}V_{nn}(x_{1})\right]}, & m\neq n\\
0, & m=n,
\end{cases}
\end{align}
and we have used the fact that $V_{mm}(x)=V_{-m,-m}(x).$ To leading
order in the disorder strength, the correlations of the new disorder
term are unaltered,
\begin{equation}
\langle V_{mn}(x)V_{mn}(x')e^{i\int_{x'}^{x}{\rm {\rm d}}x_{1}\left[\frac{1}{v_{m}}V_{mm}(x_{1})-\frac{1}{v_{n}}V_{nn}(x_{1})\right]}\rangle\simeq\gamma_{mn}\delta(x-x')+\mathcal{O}(V^{4}).
\end{equation}

\subsubsection{Born approximation}

We begin by rewriting the Hamiltonian in a BdG form
\begin{equation}
H=\int{\rm d}x\sum_{mn}\left ( \tilde{\phi}_{m}^{\dagger}(x),\tilde{\phi}_{-m}(x)\right ) \mathcal{H}_{mn}(x)\begin{pmatrix}\tilde{\phi}_{n}(x)\\
\tilde{\phi}_{-n}^{\dagger}(x)
\end{pmatrix},
\end{equation}
where $\mathcal{H}_{mn}(x)=\mathcal{H}_{mn}^{0}(x)+\mathcal{V}_{mn}(x),$
\begin{equation}
\mathcal{H}_{mn}^{0}(x)=\begin{pmatrix}-iv_{m}\partial_{x} & \Delta_{m}\\
\Delta_{m}^{\ast} & -iv_{-m}\partial_{x}
\end{pmatrix}\delta_{mn}\hspace{1em};\hspace{1em}\mathcal{V}_{mn}(x)=\begin{pmatrix}\tilde{V}_{mn}(x) & 0\\
0 & -\tilde{V}_{n,m}(x)
\end{pmatrix}e^{ik_{mn}x},\label{eq:H_BdG}
\end{equation}
where $k_{mn}=k_{{\rm F},m}-k_{{\rm F},n}.$

The diagonal elements of the disorder-averaged Green function are given
by
\begin{equation}
\begin{split}
\left\langle G_{mm}(x,x')\right\rangle  & =G_{mm}^{0}(x,x')+\int{\rm d}x_{1}G_{mm}^{0}(x,x_{1})\left\langle \mathcal{V}_{mm}(x_{1})\right\rangle G_{mm}^{0}(x_{1},x')\\
& +\int{\rm d}x_{1}\int{\rm d}x_{2}\sum_{n}G_{mm}^{0}(x,x_{1})\left\langle \mathcal{V}_{mn}(x_{1})G_{nn}^{0}(x_{1},x_{2})\mathcal{V}_{nm}(x_{2})\right\rangle G_{mm}^{0}(x_{2},x')+\dots
\end{split}
\end{equation}
Using Eqs. (\ref{eq:V_mnV_mn}-\ref{eq:V_mnV_nm}), and defining the
phase $\alpha_{mn}$ through $\gamma_{mn}=|\gamma_{mn}|\exp(i\alpha_{mn}),$
we can write
\begin{equation}
\begin{split}
\left\langle G_{mm}(x,x')\right\rangle  & =G_{mm}^{0}(x,x')+\int{\rm d}x_{1}\sum_{n}|\gamma_{mn}|G_{mm}^{0}(x,x_{1})e^{i\frac{\alpha_{mn}}{2}\tau_{z}}\tau_{z}G_{nn}^{0}(x_{1},x_{1})\tau_{z}e^{-i\frac{\alpha_{mn}}{2}\tau_{z}}G_{mm}^{0}(x_{1},x')+\dots=\\
& =\int\frac{{\rm d}q}{2\pi}e^{iq(x-x')}G_{mm}^{0}(q)\left\{ 1+\sum_{n}|\gamma_{mn}|e^{i\frac{\alpha_{mn}}{2}\tau_{z}}\tau_{z}\int\frac{{\rm d}q'}{2\pi}G_{nn}^{0}(q')\tau_{z}e^{-i\frac{\alpha_{mn}}{2}\tau_{z}}\cdot G_{mm}^{0}(q)+\dots\right\} .
\end{split}
\end{equation}
From this one extracts the self energy to leading order,
\begin{equation}
\begin{split}\Sigma_{m}(q) & =\sum_{n}|\gamma_{mn}|e^{i\frac{\alpha_{mn}}{2}\tau_{z}}\tau_{z}\int\frac{{\rm d}q'}{2\pi}G_{nn}^{0}(q')\tau_{z}e^{-i\frac{\alpha_{mn}}{2}\tau_{z}}=-\sum_{n}\frac{|\gamma_{mn}|}{2v_{n}}e^{i\frac{\alpha_{mn}}{2}\tau_{z}}\tau_{z}e^{i\arg(\Delta_{n})}\tau_{x}\tau_{z}e^{-i\frac{\alpha_{mn}}{2}\tau_{z}}\\
& =\sum_{n}\frac{|\gamma_{mn}|}{2v_{n}}e^{i\left[\arg(\Delta_{n})+\alpha_{mn}\right]\tau_{z}}\tau_{x},
\end{split}
\label{eq:self_E_SM}
\end{equation}
where we have used the expression for $G_{nn}^{0}(q')$ given in Eq.
(3) of the main text. At low energies one can construct an effective
Hamiltonian describing the $m$-th channel, $\mathcal{H}_{mm}^{{\rm eff}}(q)=\mathcal{H}_{mm}^{0}(q)+\Sigma_{m}(q)$,
where $\mathcal{H}_{mm}^{0}(q)$ is the Fourier space representation
of $\mathcal{H}_{mm}^{0}(x)$ defined in Eq. (\ref{eq:H_BdG}). This
then defines an effective pairing potential,
\begin{equation}
\Delta_{m}^{{\rm eff}}=\Delta_{m}+\frac{1}{2}\sum_{n}\frac{1}{\tau_{mn}}e^{i\left[\arg(\Delta_{n})+\alpha_{mn}\right]},\label{eq:delt_eff_SM}
\end{equation}
as appearing in Eq. (5) of the main text.

\subsubsection{Effective time-reversal symmetry}

In the case of the planar Josephson junction, the expressions for
the self energy and the effective pairing potentials, Eqs. (\ref{eq:self_E_SM})
and (\ref{eq:delt_eff_SM}), can be simplified thanks to another symmetry.
While the system breaks the usual time-reversal symmetry due to the
presence of a magnetic field, it nevertheless obeys (in the clean
limit) an anti-unitary symmetry, given by~\cite{Pientka2017topological}
\begin{equation}
[\mathcal{H}_{{\rm N}}^{0}(x,-y)]^{\ast}=\mathcal{H}_{{\rm N}}^{0}(x,y).
\end{equation}
One can therefore choose the eigenstates to obey,
\begin{equation}
\eta_{\nu,-k_x}^{s}(y)=[\eta_{\nu,k_x}^{s}(-y)]^{\ast},
\end{equation}
and together with the symmetry of Eq. (\ref{eq:ref_sym}) one has
$\eta_{\nu,k_x}^{s}(y)=\sum_{s}\sigma_{ss'}^{x}[\eta_{\nu,k_x}^{s'}(-y)]^{\ast}$.

From this one can infer that $\gamma_{mn}$ is real and positive,
\begin{align}
\gamma_{mn}^{\ast} & =\gamma\int_{-W/2}^{W/2}{\rm d}y\left[\sum_{s,s',s''}\sigma_{ss'}^{x}\sigma_{ss''}^{x}[\eta_{m,k_{{\rm F},m}}^{s'}(-y)]^{\ast}\eta_{n,k_{{\rm F},n}}^{s''}(-y)\right]^{2}=\gamma\int_{-W/2}^{W/2}{\rm d}y\left[\sum_{s}[\eta_{m,k_{{\rm F},m}}^{s}(y)]^{\ast}\eta_{n,k_{{\rm F},n}}^{s}(y)\right]^{2}=\gamma_{mn}.
\end{align}

\section{Solution by mapping to a normal disordered wire}

In the main text we have used Eq. (5) to study two special cases:
(i) the single-channel \emph{p}-wave SC, and (ii) the single-(spinful)-channel
\emph{s}-wave. While Eq (5) of the main text was derived under the
assumption of weak disorder, we here show that the results for the above
special cases are exact. Inspired by the approach of Rieder \emph{et al.}~\cite{Rieder2013reentrant}, we use a mapping
of these superconducting systems, at zero energy, to a disordered normal-metal wire, whose properties have been previously studied~\cite{Halperin1967properties,Dorokhov1982transmission,Mello1988macroscopic}.

\subsection{The (spinless) single-channel \emph{p}-wave superconductor}

The linearized Hamiltonian for single-channel \emph{p}-wave superconductor
in the presence of short-range disorder is given by
\begin{equation}
\begin{split}H_{{\rm p}}=\int{\rm d}x & \Big\{-iv\left[R^{\dagger}(x)\partial_{x}R(x)-L(x)^{\dagger}\partial_{x}L(x)\right]+\Delta\left[R^{\dagger}(x)L^{\dagger}(x)+L(x)R(x)\right]+V(x)\left[R^{\dagger}(x)R(x)+L^{\dagger}(x)L(x)\right]+\\
& +\left[V(x)e^{2ik_{{\rm F}}x}R^{\dagger}(x)L(x)+{\rm h.c.}\right]\Big\}.
\end{split}
\label{eq:p_wave_H_SM}
\end{equation}
In terms of the notation used in Eq.~(2) of the main text, $\phi_{1}(x)=R(x)$,
and $\phi_{-1}(x)=L(x)$. The above Hamiltonian can be written in
the BdG form, $H_{{\rm p}}=\frac{1}{2}\int{\rm d}x\Phi^{\dagger}(x)\mathcal{H}_{{\rm p}}(x)\Phi(x),$
\begin{equation}
\mathcal{H}_{{\rm p}}(x)=-iv\partial_{x}\sigma_{z}+V(x)\tau_{z}+V'(x)\sigma_{x}\tau_{z}-V"(x)\sigma_{y}-\Delta\tau_{y}\sigma_{y},\label{eq:p_wave_H_BdG}
\end{equation}
where here $\Phi^{\dagger}(x)=[R^{\dagger}(x),R(x),L^{\dagger}(x),L(x)]$,
and $V'(x)=V(x)\cos(2k_{{\rm F}}x)$, $V"(x)=V(x)\sin(2k_{{\rm F}}x)$.
The disorder potential $V(x)$ is described by the correlations $\langle V(x)V(x')\rangle=\gamma_{\rm p}\delta(x-x')$.

The localization length, at a given energy, can be obtained from the
transfer matrix, $M(x,\varepsilon)$, defined as the $4\times4$ matrix
obeying
\begin{equation}
\Phi(x,\varepsilon)=M(x,\varepsilon)\cdot\Phi(0,\varepsilon),\label{eq:trans_mat_def}
\end{equation}
where $\Phi(x,\varepsilon)\equiv\int{\rm d}t\Phi(x,t)\exp\left(-i\varepsilon t\right)$,
and propagation in time is according to $H_{{\rm p}}$. The localization length is related to the transfer matrix through the eigenvalues of $M^\dagger M$, which in the localized phase take the form $\exp(\pm2\lambda_i L)$ when $L\to\infty$, where $\{\lambda_i\}$ are the so-called Lyapunov exponents~\cite{Beenakker1997Random,Evers2008Anderson}. The localization length is then determined by the slowest decaying exponent, $\xi=1/\max\{\lambda_i\}$

Writing the Hamiltonian, Eq.~\eqref{eq:p_wave_H_BdG}, as $\mathcal{H}_{\rm p}=-iv\sigma_z(\partial_x+\mathcal{H}_1)$, the Schr\"odinger equation for $\Phi(x,\varepsilon)$ takes the form
$\partial_x\Phi(x,\varepsilon)=(i\sigma_z\varepsilon/v-\mathcal{H}_1)\Phi(x,\varepsilon)$, which is solved by $\Phi(x,\varepsilon)=\mathcal{T}_{x}\exp[i\sigma_z\varepsilon x/v - \int_0^x {\rm d}x'\mathcal{H}_1(x')]\Phi(0,\varepsilon)$, where $\mathcal{T}_{x}$ is the path ordering operator. Namely the zero-energy transfer matrix from one side of the system to the other is given by
\begin{equation}
M(L,\varepsilon=0)=\mathcal{T}_{x}\exp\left\{ \frac{1}{v}\int_{0}^{L}{\rm d}x\left[-iV(x)\sigma_{z}\tau_{z}+V'(x)\sigma_{y}\tau_{z}+V"(x)\sigma_{x}+\Delta\sigma_{x}\tau_{y}\right]\right\} .\label{eq:p_wave_trans_mat}
\end{equation}
The last term in the exponent, $\Delta\tau_{y}\sigma_{x}$, commutes with all other terms. Therefore, the transfer matrix decomposes into two $2\times 2$ blocks, $M_\pm$, where the $\pm$ refers to the eigenvalue of $\tau_y\sigma_x$. These blocks are given by, 
\begin{equation}\label{eq:M_from_M_N}
M_\pm(L,\varepsilon=0)=M_{\rm N}(L,\varepsilon=0)e^{\pm\Delta L/v}
\end{equation}
where $M_{\rm N}(L,\varepsilon)$ is the transfer matrix for a single-channel normal wire of linear dispersion.

The problem of a normal disordered wire has been solved elsewhere~\cite{Halperin1967properties,Dorokhov1982transmission,Mello1988macroscopic}, and the resulting eigenvalues of $M_{{\rm N}}^{\dagger}(L,0)M_{{\rm N}}(L,0)$ read $e^{\pm2\lambda_{{\rm N}}L},$ where $\langle\lambda_{{\rm N}}\rangle\underset{L\to\infty}{\to}1/2l$
and its variance goes to zero. From Eq.~\eqref{eq:M_from_M_N} we then conclude that the four eigenvalues of $M^\dagger(L,0)M(L,0)$ are given by $e^{\pm2\left(\lambda_{{\rm N}}\pm\Delta/v\right)L},$
which means that the zero-energy localization length for the \emph{p}-wave SC reads
\begin{equation}
\frac{1}{\xi_{{\rm p}}}=\left|\frac{1}{\xi_{{\rm p}}^{0}}-\frac{1}{2l}\right|,\label{eq:xi_p_wave}
\end{equation}
where $\xi_{{\rm p}}^{0}=v/\Delta$, in accordance with the result shown below Eq. (6) of
the main text. While that result was obtained from a perturbative weak-disorder
treatment, the calculation leading to Eq. (\ref{eq:xi_p_wave}) is exact (within the linearized
model).

\subsection{The (spinful) single-channel \emph{s}-wave superconductor}

We now move on to a single spinful channel \emph{s}-wave SC.\emph{
}The linearized Hamiltonian for such a system is given by
\begin{equation}
\begin{split}H_{{\rm s}}=\int{\rm d}x\Big( & \sum_{s=\uparrow\downarrow}\left\{ -iv\left[R_{s}^{\dagger}(x)\partial_{x}R_{s}(x)-L_{s}^{\dagger}(x)\partial_{x}L_{s}(x)\right]+V(x)\left[R_{s}^{\dagger}(x)R_{s}(x)+L_{s}^{\dagger}(x)L_{s}(x)\right]+\left[V(x)R_{s}^{\dagger}(x)L_{s}(x)+{\rm h.c.}\right]\right\} \\
& +\Delta\left[R_{\uparrow}^{\dagger}(x)L_{\downarrow}^{\dagger}(x)+L_{\uparrow}^{\dagger}(x)R_{\downarrow}^{\dagger}(x)+{\rm h.c.}\right]\Big).
\end{split}
\label{eq:s_wave_H}
\end{equation}
We can write it in the BdG form $H_{{\rm \text{s}}}=\int{\rm d}x\Phi^{\dagger}(x)\mathcal{H}_{{\rm s}}(x)\Phi(x),$
\begin{align}
\mathcal{H}_{{\rm s}}(x) & =-iv\partial_{x}\sigma_{z}+V(x)\tau_{z}+V'(x)\sigma_{x}\tau_{z}-V"(x)\sigma_{y}+\Delta\tau_{x}\sigma_{x}.\label{eq:s_wave_H_BdG}
\end{align}
where this time $\Phi^{\dagger}(x)=[R_{\uparrow}^{\dagger}(x),R_{\downarrow}(x),L_{\uparrow}^{\dagger}(x),L_{\downarrow}(x)]$. This Hamiltonian resembles the $p$-wave BdG Hamiltonian of Eq.~(\ref{eq:p_wave_H_BdG}), except for the matrix structure of the pairing term, $\Delta$. This difference comes from the fact that in the $p$-wave case, the pairing potential switches sign when going from positive to negative momenta. Notice that even though the \emph{s}-wave SC is spinful, we could defined the BdG matrix, $\mathcal{H}_{{\rm s}}$, such that it would have the same size as $\mathcal{H}_{{\rm p}}$. This is possible only because the disorder term in Eq.~\eqref{eq:s_wave_H} does not mixes opposite spins.

We can obtain an expression for the transfer matrix in exactly the same way as we did above for the $p$-wave case [see Eq.~\eqref{eq:p_wave_trans_mat}]. This results in
\begin{equation}
M(L,\varepsilon=0)=\mathcal{T}_{x}\exp\left\{ \frac{1}{v}\int_{0}^{L}{\rm d}x\left[-iV(x)\sigma_{z}\tau_{z}+V'(x)\sigma_{y}\tau_{z}+V"(x)\sigma_{x}+\Delta\sigma_{y}\tau_{x}\right]\right\}.
\end{equation}
Unlike in the $p$-wave case, this time the pairing term, $\Delta\tau_{x}\sigma_{y}$, does \emph{not} commute with the rest of the terms in the exponent. Nevertheless, all terms in the exponent still commute with $\tau_{y}\sigma_{x}$. We can therefore decompose $M(L,0)$ into two blocks by going to the basis which diagonalizes $\tau_{y}\sigma_{x}$. This is done by $\tilde{M}(L,0)=\mathcal{U}^{\dagger}M(L,0)\mathcal{U}=M_+\oplus M_-$, where $\mathcal{U}=\frac{1}{2}\left[1+\tau_{z}+\sigma_{x}(1-\tau_{z})\right]$$e^{i\frac{\pi}{4}\tau_{x}}$, and where
\begin{equation}\label{eq:M_pm_tilde}
\tilde{M}_\pm =\mathcal{T}_{x}\exp\left\{ \frac{1}{v}\int_{0}^{L}{\rm d}x\left[-iV(x)\sigma_{z}+V'(x)\sigma_{y}+V"(x)\sigma_{x}\pm\Delta\sigma_{z}\right]\right\}.
\end{equation}

In the absence of $\Delta$, the matrices $\tilde{M}_\pm$ both correspond again to the transfer matrix of single-channel normal disordered wire (with linear dispersion). Importantly, we notice that $\Delta$ enters in Eq.~\eqref{eq:M_pm_tilde} as an imaginary energy, $V(x)\to V(x) \pm i\Delta$, namely
\begin{equation}
\tilde{M}_\pm = M_{{\rm N}}(L,\varepsilon=\pm i\Delta).
\end{equation}
Namely, the zero-energy \emph{s}-wave transfer matrix is mapped
to two copies of a normal disordered wire at finite energy, with the analytic continuation, $\varepsilon\to\pm i\Delta$.

To perform the analytic continuation, we first use the Friedel sum
rule. For the case of a single-channel normal wire, it relates the
reflection amplitude for a system with \emph{open boundary conditions,
	$r_{{\rm obc}}(L,\varepsilon)=e^{i\varphi(\varepsilon,L)}$,} to the
density of states per unit length, $\nu(\varepsilon)$, through
\begin{equation}
\nu(\varepsilon)=\lim_{L\to\infty}\frac{1}{2\pi L}\frac{\partial\varphi(\varepsilon,L)}{\partial\varepsilon},
\end{equation}
where $r_{{\rm obc}}(L,\varepsilon)$ is the reflection for an electron
incident at $x=0$, with a boundary condition $\Phi(x=L)=0$. For
the linearized model of the disordered wire, the density of states
(in the thermodynamic limit) is constant, $\nu=1/2\pi v$,
yielding
\begin{equation}
\varphi(\varepsilon,L)=\varphi_{0}(L)+\varepsilon L/v,\label{eq:phi}
\end{equation}

The above reflection amplitude, $r_{{\rm obc}}$, is related to the
transfer matrix through
\begin{equation}\label{eq:r_obc_to_M_N}
\begin{pmatrix}r_{{\rm obc}}\\
1
\end{pmatrix}=M_{{\rm N}}(L,\varepsilon)\begin{pmatrix}1\\
1
\end{pmatrix}.
\end{equation}
We write the transfer matrix of the normal wire using its polar decomposition~\cite{Beenakker1997Random,Evers2008Anderson},
\begin{equation}
M_{{\rm N}}(L,\varepsilon)=\begin{pmatrix}e^{i\alpha} & 0\\
0 & e^{-i\alpha}
\end{pmatrix}\begin{pmatrix}\cosh(\mu) & \sinh(\mu)\\
\sinh(\mu) & \cosh(\mu)
\end{pmatrix}\begin{pmatrix}e^{i\beta} & 0\\
0 & e^{-i\beta}
\end{pmatrix},\label{eq:M_N_pol_dec}
\end{equation}
where the parameter $\mu$ is related to the Lyapunov exponent by $\mu=\lambda_{N}L$, when $L\to\infty$. From Eqs.~\eqref{eq:r_obc_to_M_N} and \eqref{eq:M_N_pol_dec} we then conclude that $r_{{\rm obc}}=e^{2i\alpha}$, namely $\alpha(L,\varepsilon)\underset{L\to\infty}{\to}\alpha_{0}(L)+\varepsilon L/2v$.
Applying the same arguments for an electron incident towards the left at $x=L$, with open boundary conditions at $x=0$, one concludes that
$\beta(L,\varepsilon)\underset{L\to\infty}{\to}\beta_{0}(L)+\varepsilon L/2v$.

We can now perform the analytic continuation,
\begin{equation}
\begin{split}M_{{\rm N}}(L,\varepsilon\to i\Delta)= & \begin{pmatrix}e^{i\alpha_{0}} & 0\\
0 & e^{-i\alpha_{0}}
\end{pmatrix}\begin{pmatrix}e^{-\Delta L/v}\cosh(L/2l) & \sinh(L/2l)\\
\sinh(L/2l) & e^{\Delta L/v}\cosh(L/2l)
\end{pmatrix}\begin{pmatrix}e^{i\beta_{0}} & 0\\
0 & e^{-i\beta_{0}}
\end{pmatrix}.\end{split}
\end{equation}
Finally, one computes the eigenvalues of $M_{{\rm N}}(L,i\Delta)[M_{{\rm N}}(L,i\Delta)]^{\dagger},$
which are given by
\[
e^{\pm2\cosh^{-1}[\cosh(L/2l)\cosh(\Delta L/v)]}\underset{L\to\infty}{\longrightarrow}e^{\pm2(1/2l+\Delta/v)L}.
\]
A similar results is obtained for $M_{{\rm N}}(L,-i\Delta)[M_{{\rm N}}(L,-i\Delta)]^{\dagger}$,
so that altogether we get
\begin{equation}
\frac{1}{\xi_{{\rm s}}}=\frac{1}{\xi_{{\rm s}}^{0}}+\frac{1}{2l},
\end{equation}
in accordance with the result of Eq.~(7) of the main text.

\end{widetext}
	
\end{document}